\documentclass[preprint,12pt]{elsarticle}

\usepackage{graphicx}
\usepackage{amssymb}

\usepackage{lineno}

\usepackage{amsmath}
\usepackage{amssymb}
\usepackage{amsfonts}
\usepackage{multirow}
\usepackage{amsthm}
\RequirePackage[OT1]{fontenc}
\RequirePackage{amsthm,amsmath,amsfonts}
\RequirePackage{natbib}
\RequirePackage[colorlinks,citecolor=blue,urlcolor=blue]{hyperref}

\newtheorem{proposition}{Proposition}
\theoremstyle{definition}

\newcommand{\lam}{\ensuremath{\lambda(\boldsymbol{s_i},t)}}

\newcommand{\si}{\ensuremath{\boldsymbol{s_i}}}

\journal{Spatial Statistics}

\begin{document}

\begin{frontmatter}


\title{A CLASS OF SPATIALLY CORRELATED SELF-EXCITING STATISTICAL MODELS}



\author{Nicholas J. Clark}

\address{United States Military Academy}

\author{Philip M. Dixon}
\address{Iowa State University}
\begin{abstract}
 The statistical modeling of multivariate count data observed on a space-time lattice has generally focused on using a hierarchical modeling approach where space-time correlation structure is placed on a continuous, latent, process.  The count distribution is then assumed to be conditionally independent given the latent process.  However, in many real-world applications, especially in the modeling of criminal or terrorism data, the conditional independence between the count distributions is inappropriate.  In this manuscript we propose a class of models that capture spatial variation and also account for the possibility of data model dependence. The resulting model allows both data model dependence, or self-excitation, as well as spatial dependence in a latent structure.  We demonstrate how second-order properties can be used to characterize the spatio- temporal process and how misspecificaiton of error may inflate self-excitation in a model.  Finally, we give an algorithm for efficient Bayesian inference for the model demonstrating its use in capturing the spatio-temporal structure of burglaries in Chicago from 2010-2015.
\end{abstract}


\end{frontmatter}


\section{Introduction}
\label{S:1}

The modeling of count data where each observation takes place on a space-time lattice arises in multiple disciplines.  In the disease literature, the number of infected patients is often aggregated over geographic areas and discrete times to protect the confidentiality of patients \cite{fefferman2005confidentiality}.  In the modeling of terrorism or criminal acts, as we consider in this manuscript, data is often presented aggregated over time and space for security reasons \cite{python2019bayesian,clark2018modeling}.  Even for spatial continuous and temporally continuous data, the analysis is often performed aggregated over fixed spatial and temporal domains as a matter of convenience \cite{brantingham2009crime, bernasco2011robberies}.  The challenge is how to appropriately model the relationship between observations.  Assumptions on either the spatial relationship or the temporal relationship between observations are necessary if any statistical analysis is to be performed.  In this paper we present a novel approach for structuring space-time dependency for count data through a combination of spatial dependence in a latent process model and temporal dependence in a data model.  To do this, we derive a stochastic difference equation for the intensity of the space-time process rather than placing structure on a latent space-time process as is commonly proposed, e.g. Chapter 6 of \cite{cressie2015statistics}.

In the spatial statistics literature, an early attempt at structuring spatial relationships for count data was made in \cite{besag1974spatial} where the data model distribution was conditionally specified given a fixed spatial neighborhood.  However, as shown in \cite{besag1974spatial}, this results in a statistical model that only allows negative correlation.  More recently, \cite{kaiser1997modeling} demonstrated how modifications could be made to the statistical model that allowed both negative and positive correlation.  A similar methodology was employed in \cite{augustin2006using} to address the spatial dynamics of seed count data in agricultural models.  The critical assumption in these classes of models is that the distribution of the observed counts can be conditionally specified from the observed counts at spatial neighbors, a Markov assumption in space.

Advances in computation and Bayesian inference have also allowed for modeling through spatial hierarchical models similar to the Poisson log-Normal approach of \cite{aitchison1989multivariate}. Letting $\boldsymbol{s_i}$ be a discrete spatial location and $t$ be discrete time, spatio-temporal dependence can be introduced by assuming the existence of a latent Gaussian process, $X(\boldsymbol{s_i},t)\sim \mbox{Gau}(\boldsymbol{\alpha},\Sigma(\theta))$ that has a spatio-temporal structure characterized by $\Sigma(\theta)$.  The data model is then assumed to be independent given the latent state, for instance 

\begin{equation}
    Y(\boldsymbol{s_i},t) \sim \mbox{Pois}\left(\exp(X(\boldsymbol{s_i},t)\right)
\end{equation}. The idea was extended to incorporate spatial dynamics in \cite{besag1991bayesian}.  Here a spatial Markov assumption was still made, but it was made in a latent, unobserved, continuous process.  The spatial observations were then assumed to be independent given the latent process.  This idea was also used in \cite{wolpert1998poisson}, who used a Poisson-gamma model with spatial dependence in the latent, gamma, structure.

The concept of allowing the spatial dependence to exist only in a latent field overcomes the difficulty of only negative spatial correlation that arises in the auto-Poisson model of \cite{besag1974spatial}.  Although this form of modeling only allows for limited dependence in the data as demonstrated in \cite{aitchison1989multivariate}, it has become commonplace in literature.  For example in \cite{goicoa2016age} mortality rates are studied using latent conditional effects for space, time, and age. However, these approaches assume that the dependence in the data is due to a latent, unobserved process which does not capture repeat victimization that is believed to exist in crime and violence.   Repeat victimization, explained for example in \cite{polvi1990repeat}, is the belief that an observed crime or violent act increases the likelihood of a future crime occurring at that exact same spot or against the exact same person and can be modeled assuming a data model dependence or as an observation driven process.

While count data in the spatial statistics literature has predominately been addressed through structure in a latent process, in the time series literature it has evolved quite differently.  For example, the INGARCH model of \cite{ferland2006integer} and \cite{heinen2003modelling} is a time series model for counts where the data model is Poisson with expectation that is a function of both previous counts and previous expectations.  Specifically, if we let ${Z_t}$ be a time series of counts and $\mathcal{F}_t$ be the $\sigma$-field generated by ${Z_0,...,Z_t,\lambda_0}$, the INGARCH$(p,q)$ model is

\begin{equation}
	Z_t|\mathcal{F}_{t-1}\sim \mbox{Pois}(\lambda_t),\quad \lambda_t=d + \sum_{i=1}^p a_i \lambda_{t-i}+\sum_{j=1}^q b_j Z_{t-j}\label{eq:INGARCH}
\end{equation}
This results in a time series model that is a function of both the data model and a deterministic process model.  \cite{ferland2006integer} demonstrated how the INGARCH(1,1), given as $\lambda(s_i,t)=d+\kappa \lambda(s_i,t-1)+\eta Y(s_i,t-1)$, is analogous to an ARMA(1,1) for counts where $d$ is the baseline intensity, $a_1 = \kappa$ is synonymous with the auto-regressive parameter and $b_1 = \eta$ is similar to the moving average parameter. In \cite{fokianos2009poisson} it was shown that a perturbed INGARCH(1,1) model was geometrically ergodic giving a unique stationary distribution and asymptotic normality of the roots of the likelihood equations.  The stationary distribution of the INGARCH(1,1) process is also equivalent to a stochastic process given in \cite{hawkes1971spectra}, often called a self-exciting point process, where

\begin{align}
	Y(t)|\lambda_t& \sim \mbox{Pois}(\lambda(t))\label{eq:Hawkes}\\
	\lambda(t) & = \nu(t) + \int_{0}^{t} g (t-u) N(ds)\nonumber,
\end{align}
when the process is sampled at discrete times and $g(t-u)=\eta \exp(-\alpha (t-t_i))$.  Note here we use $\eta$ to represent the excitation term in the model similar to what is done above but the parameters have different interpretation. In \eqref{eq:Hawkes}, $\nu(t)$ is often referred to as the background intensity and $g(t-u)$ is referred to as the excitation function, see \cite{laub2015hawkes}.

While the INGARCH model was motivated to model univariate time series data, there has been some recent effort to apply it to multivariate count data.  \cite{heinen2007multivariate} used copulas to model the contemporaneous correlation.  However there are issues with using copulas for count data and it is generally less reliable and identifiable than the use of copulas for continuous data, as explained in \cite{genest2007primer}.  \cite{liu2012some} allows for a spatial lag dependency through treating $\lambda_t$ as a vector and replacing $a_i$ and $b_j$ with a series of matrices.  The author then allows for contemporaneous correlation through the bivariate Poisson.  These models, though, do not naturally extend previous spatial models to the INGARCH class nor do they capture how criminologists and others believe crime actually evolves over space and time.  Furthermore, as we will show, the variance to mean ratio for the INGARCH process dictates the range for the allowable autocorrelation limiting its practical use.

In this manuscript, we introduce a class of Self-Exciting Spatio-Temporal models for count data we refer to as Spatially Correlated Self-Exciting or SCSE models that capture both data model dependence as well as dependence in a latent spatial process. We will demonstrate how these models arise from natural assumptions on how crime and violence evolves over space and time, how they retain the same stationarity properties as the INGARCH model and how they can be differentiated through assessment of second order characteristics.  The SCSE class also allows a much wider range of second order properties affording the modeler more flexibility in describing the autocorrelation and variance to mean ratio of the data.  We will further show how to conduct inference and model assessment to differentiate between models within this class using burglaries in Chicago as an example.

\section{General Model}

Throughout this manuscript we use the following definitions unless specified.  $\textit{\si}$ is used to denote a vector of spatial lattice locations that remain fixed in time, $t$ is a discrete time period.  We allow $N|\si|$ to denote the spatial neighborhood of lattice location $\si$.  $Y(\si,t)$ is the observed process at spatial location $\si$ and time $t$ and $X(\si)$ is the unobserved latent process that is unique to spatial location $\si$.

The complete model is a stochastic difference equation operating directly on the intensity function $\lam$:

\begin{align}
	& Y(\boldsymbol{s_i},t)|\lambda(\boldsymbol{s_i},t) \sim \mbox{Pois}(\lambda(\boldsymbol{s_i},t)) \label{eq:timeseries2} \\
	& E[Y(\boldsymbol{s_i},t)|\lambda(\boldsymbol{s_i},t)]=\lambda(\boldsymbol{s_i},t)\nonumber\\
	& \boldsymbol{\lambda}_t = \exp(\boldsymbol{X}+\boldsymbol{\epsilon}_t)+\eta \boldsymbol{Y}_{t-1}+\kappa \boldsymbol{\lambda}_{t-1}\nonumber\\
	& \boldsymbol{X} \sim \mbox{Gau} (\boldsymbol{\alpha},(I_{{n},{n}}-\zeta\mathbf{C})^{-1}\sigma^2),\nonumber\\
	& \boldsymbol{\epsilon}_t \sim \mbox{Gau} (\boldsymbol{0},I \sigma_{\epsilon}^2),\nonumber
\end{align}

where $\boldsymbol{\lambda_t}=\left(\lambda(s_1,t),\lambda(s_2,t),\cdots,\lambda(s_{n},t)\right)^T$. 

In \eqref{eq:timeseries2}, large scale spatial structure is accounted for in the latent process $X(\si)$ through spatial regression parameter $\alpha$ whereas small scale spatial structure is accounted for through conditionally specifying $X(\si)$. Here we condition on the spatially adjacent neighbors

\begin{align}
	& X(\si)|x\left(s_j | s_j \in N|\si|\right)\sim N(\mu(\si),\sigma^2) \label{eq:Latent Dependency}\\
	& \mu(\boldsymbol{s_i}) = \alpha(\si)+ \zeta \sum_{s_j \in N|\si|} \{x(s_j)-\alpha(s_j)\} \nonumber.
\end{align}

 Letting $\mathbf{C}$ be a matrix with entry in the $i,j$ have entries of 1 if locations $\boldsymbol{s}_j$ and $\si$ are neighbors, Brook's Lemma yields a joint density for $\boldsymbol{X}$ as:
 
\begin{equation}
    \boldsymbol{X} \sim \mbox{Gau}\textit{} (\boldsymbol{\alpha},(I_{{n},{n}}-\zeta\mathbf{C})^{-1}\sigma^2),
\end{equation}
as given in \eqref{eq:timeseries2}. The $\zeta$ parameter, in general, controls the amount of spatial dependence in the model not captured by covariates in $\boldsymbol{\alpha}$.
The spatial model given in \eqref{eq:Latent Dependency} is commonly referred to as a Conditional Auto-Regressive (CAR) model (see e.g. Section 4.2.5 of \cite{cressie2015statistics}). To further allow variation in the spatio-temporal process we can add additional space-time noise letting $\epsilon(\si,t)\stackrel{iid}{\sim}N(0,\sigma_{\epsilon})$.

Restricting $\eta$ and $\kappa$ to be non-negative, it is clear that $\boldsymbol{\lambda_t}$ is a Markov chain on state space $(\mathbb{R}^{+})^{n}$.  Letting $\chi = \kappa - 1$ at each spatial location, $\boldsymbol{s_i}$, the change in expectation, $\Delta \lambda(\boldsymbol{s_i},t)=\lambda(\boldsymbol{s_i},t)-\lambda(\boldsymbol{s_i},t-1)$, is

\begin{equation}
	\Delta \lambda(\boldsymbol{s_i},t)=\exp\left(X(\si)+\epsilon(\si,t)\right)-\chi \lambda(\boldsymbol{s_i},t-1)+\eta Y(\boldsymbol{s_i},t-1) \label{eq:SCSEDDS}.
\end{equation}
If, for example, this process was used to model the number of violent events that occur at space-time location ($\boldsymbol{s_i},t$) we can, similar to the mathematical model for crime given in \cite{short2008statistical}, think of $\lambda(\boldsymbol{s_i},t)$ as representing the tension at location $\boldsymbol{s_i}$ at time $t$.  The change in tension is then a function of large-scale spatially varying exogenous factors, $\boldsymbol{\alpha}$.  The $\eta$ term captures the expected change  that is due to repeat or near-repeat actions, a characteristic of violence that has been shown to exist in the social science literature, see e.g. \cite{polvi1990repeat} and \cite{pease1998repeat} whereas $\kappa$ captures decay in tension absent repeat victimization.  The unobservable $\exp\left(X(\boldsymbol{s_i})+\epsilon(\si,t)\right)$ term, then, can be thought of as the baseline change in tension in the absence of any endogenous factors in the model. 

A complete treatment of the relationship between the SCSE model and self-exciting processes is given in \ref{sec:SCSE and Self-Exciting Process}.

\subsection{Conventional Spatio-Temporal Statistical Models and Related Models}
The most common methodology for capturing spatio-temporal structure in a statistical model is to use a multivariate Poisson-log Gaussian distribution and place spatio-temporal structure on the covariance matrix of the log Gaussian.  A key assumption in this approach is that the data model, $Y(\si,t)$ can be defined conditionally on a process model, $X(\si,t)$.  The process model, then, is a function of both observable spatial or temporal covariates as well as unobservable latent spatial errors.  This technique has been used to model the spatio-temporal structure in criminal behavior \cite{hu2018urban}.  Here, the authors modeled the number of crimes at location $\si$ and time $t$ as $Y(\si,t)\sim \mbox{Binom}(n(\si,t),X(\si,t))$.  The assumption that is made is that the number of crimes that occur at location $\si$ and time $t$ are conditionally independent from the other locations given $X(\si,t)$.  Further structure was then put on $\mbox{logit} (X(\si,t))=\alpha+\beta x(\si)+u(\si)+\gamma t+\delta(\si) t+e(\si)$.  Here $\beta x(\si)$ captured the large scale spatial covariates, $u(\si)$ was allowed to be spatially unstructured random effect and $e(\si)$ indicated a spatially structured random effect.  The model further captured temporal trends through the covariate $\gamma$ and a fixed spatio-temporal interaction term $\delta (\si)$ was also considered.

Formulations such as above are generally used due to both flexibility and convenience.  However, as opposed to the SCSE model, the choice of a latent spatial structure and temporal trend does not offer straight forward interpretability.  Alternatively, a hierarchical approach similar that recommended in \cite{cressie2015statistics} could be used.  Here it is proposed to use dynamical spatio-temporal statistical modeling.  Here $X(\si,t)$ is conditionally modeled given either $X(\boldsymbol{s_j},t)$ or $X(\si,t-1)$, etc.  In other words, the latent process evolves over space and time.  Care must be exercised in this approach to ensure that the resulting covariance matrix for the multivariate Gaussian $\boldsymbol{Y}$ is positive semi-definite. However, again, this is not as interpretable as the SCSE model and, as we show in Section\ref{sec:Chicago}, may not be able to replicate the second order properties of the data.

While the use of the multivariate Poisson-log Gaussian has dominated literature, it comes with the perhaps undesirable property that manipulation of the temporal correlation comes at a cost of potentially severely impacting overdispersion, see e.g. Fig 1 of \cite{aitchison1989multivariate}.  Furthermore, the interpretation of conditional independence given a latent log Gaussian state is potentially confusing for practitioners and does not reflect the belief among mathematical criminologists that there is a positive correlation in the data model itself.
 
Though the SCSE process uniquely combines common models from the spatial literature an the time series literature, there are a few other notable models that are similar.  In \cite{martinez2008autoregressive} the count of diseases was modeled on a space time lattice.  This model assumed that the number of infected individuals was conditionally Poisson where the natural parameter was structured to be a Log INGARCH (1,0) combined with a latent process model.  The latent process model was then conditionally specified similar to a spatial conditional auto-regressive CAR model and a temporal auto-regressive CAR model,  

\begin{align}
	& Y(\boldsymbol{s_i},t)\sim \mbox{Pois}(\exp(r(\boldsymbol{s_i},t))\\
	& r(\boldsymbol{s_i},t) = \mu+\alpha_t + \rho \left(r(\boldsymbol{s_i},t-1) -\alpha_{t-1}-\mu\right) + \theta(\boldsymbol{s_i},t)+\epsilon(\boldsymbol{s_i},t)\\
	& \alpha \sim \mbox{AR}(1)
	\quad \theta \sim \mbox{CAR} \quad \epsilon \sim \mbox{Gau}(0,\sigma^2).
\end{align}
Here the log-relative risk at location $\boldsymbol{s_i}$ and time $t$, $r(\boldsymbol{s_i},t)$ is a linear function of a latent Gaussian conditionally autoregressive term (CAR) in space, a latent Auto-regressive (AR) term in time, as well as a function of the previous log relative risk, $r(\boldsymbol{s_i},t-1)$.

In \cite{mohler2013modeling}, a discretized Cox-Hawkes model was presented that is an INGARCH(0,q) combined with a latent log-Gaussian process where the Gaussian process follows an Auto-Regressive (1) process,

\begin{align}
	& Z_t\sim \mbox{Pois}(\lambda_t),\quad \lambda_t=\exp(Y_t)+\sum_{j<t} \eta \kappa^{t-j} Z_{t-j} \\
	& Y_t|Y_{t-1}=y_{t-1} \sim \mbox{Gau} (\phi y_{t-1},\sigma^2).
\end{align}

The SCSE model, the \cite{martinez2008autoregressive} model, and \cite{mohler2013modeling} the model are each justified through the assumed existence of two separate processes that impact the expectation, or the log-expectation.  \cite{mohler2013modeling} used a temporal AR(1) latent process that impacts the expectation as well as a `self-exciting' proces.  Similarly, \cite{martinez2008autoregressive} used a latent Spatio-temporal CAR process combined with a data driven process that impact the log-expectation of the Poisson. 

The addition of covariates in a self-exciting spatial point process was used to model crime in \cite{reinhart2018self}.  There, the authors considered a more general marked self-exciting point process,

\begin{align}
\lambda(\boldsymbol{s_i},t)=\exp(\beta X(\boldsymbol{s_i}))+\sum_{i:t_i<t} g(\boldsymbol{s}-\boldsymbol{s_i},t-t_i,M_i),
\end{align}
where $g$ was a Gaussian kernel and $M_i$ was the type of criminal activity.

\section{SCSE Model Properties}
As is common for spatio-temporal models we concern ourselves with second order spatial and temporal properties for the SCSE process which we derive below.  We begin by first noting that the SCSE process is geometrically ergodic and converges to a unique stationary distribution.  To demonstrate this, we note similarities between \eqref{eq:timeseries2} and the Poisson auto-regressive model of \cite{fokianos2009poisson} however \eqref{eq:timeseries2} does not require a perturbation as the $\exp\left(X(\si)+\epsilon(\si,t)\right)$ term ensures that the support of \lam is $\mathbb{R}^+$

\begin{proposition} \label{Prop 1}
	Under the parameter space restriction, $\eta,\kappa\geq0$ and $\eta+\kappa<1$, the SCSE process is geometrically ergodic and admits a unique stationary distributions that has finite first two moments.
\end{proposition}

A complete proof of Proposition 1 relies on Markov chain theory and is given in Supplementary Materials.  As a result of proposition 1, we can use the stationary distribution to derive first and second order properties for the SCSE  model. 

\subsection{First Order Properties}
To derive the expectation for data from the Self-Exciting Poisson CAR model, we first note that $\exp(\boldsymbol{Y}+\boldsymbol{\epsilon}_t)$ has a multivariate log-normal distribution.  We define $\Sigma_{i,i}$ as the $i,i$th element of the covariance matrix of $\boldsymbol{Y}+\boldsymbol{\epsilon}_t$.  Therefore, as the natural parameter is linked exponentially with the linear predictor, using properties of the Poisson distribution, we have 

\begin{align}
	E\left[Y(\boldsymbol{s_i},t)\right] & = E\left[E\left[Y(\boldsymbol{s_i},t)|\lambda(\boldsymbol{s_i},t)\right]\right] \nonumber\\
	&= E\left[\lambda(\boldsymbol{s_i},t)\right]= E\left[\exp(X(\si)+\epsilon(\si,t)\right]+\eta E\left[Y(\boldsymbol{s_i},t-1)\right] +\kappa E\left[\lambda(\boldsymbol{s_i},t-1)\right] \nonumber\\
	& = \exp\left(\alpha(\si)+\frac{\Sigma_{1,1}}{2}\right)+\eta E\left[\lambda(\boldsymbol{s_i},t-1)\right] +\kappa E\left[\lambda(\boldsymbol{s_i},t-1)\right] \label{eq:Expectation},
\end{align}
which, at stationarity, yields, $E\left[Y(\boldsymbol{s_i},t)\right]=\frac{1}{1-\eta-\kappa}\exp(\alpha(\si)+\frac{\Sigma_{i,i}}{2})$.  The existence of self-excitation within the data model, or $\eta>0$, increases the marginal expectation for the data model.  

\subsection{Second Order Properties}
The SCSE model allows for flexible modeling of variances.  In particular, overdispersion can be modeled independently of temporal autocorrelation.  This key property differentiates the model from the Poisson-log normal  

\subsubsection{Variance}

Under the conditions in Proposition 1 we have second order temporal stationarity and subsequently,

\begin{equation}
	\mbox{Var }(Y(\boldsymbol{s_i},t))=\frac{1}{1-(\kappa+\eta)^2} \mbox{Var }\left[\exp(U(\boldsymbol{s_i},t))\right]+\frac{1-\kappa^2-2\kappa \eta}{1-(\kappa+\eta)^2}E\left[Y(\boldsymbol{s_i},t)\right].
\end{equation}

Therefore, the SCSE process allows for the modeling of overdispersion.  Furthermore, as we will show below, overdispersion can be accounted for while only minimally impacting the range of possible autocorrelation, which is a critical difference between this process and the Poisson Log-Normal formulation where these properties are closely linked.

\subsubsection{Temporal Covariance}

To see the impact of adjusting the mean to variance ratio on the temporal covariance we find the lag-one autocorrelation by relying on the second order stationarity implicit in Proposition 1.  As derived in Appendix B the autocovariance under the SCSE model is
\begin{equation}
	\mbox{Cov}\left[Y(\boldsymbol{s_i},t),Y(\boldsymbol{s_i},t+1)\right]=\left(\eta+\kappa\right)\mbox{Var} \left[Y(\boldsymbol{s_i},t)\right]-\kappa E[Y(\boldsymbol{s_i},t)] \label{eq:Autocov}.
\end{equation}
In particular, if $\kappa=0$ in \eqref{eq:timeseries2}, the lag-one autocorrelation for the process is $\eta$ and in general, the lag-$h$ autocorrelation is $\eta^h$.  

The significance of this is that it allows a great deal of flexibility in capturing second order properties of the data.  The SCSE process with $\kappa=0$, for example, has a lag-one auto correlation of $\eta$, and a variance to mean ratio of $\frac{\mbox{Var} \left[\exp(U(\boldsymbol{s_i},t))
	\right]}{(1-\eta) \mbox{E}\left[\exp(U(\boldsymbol{s_i},t))\right]} + \frac{1}{1-\eta^2}$.  Therefore, through manipulating the $\frac{ \mbox{Var} \left[\exp(U(\boldsymbol{s_i},t))\right]}{ \mbox{E}\left[\exp(U(\boldsymbol{s_i},t))\right]}$ we can manipulate the variance to mean ratio through only minimally impacting the autocorrelation.  If, for example, we desire data that has a variance to mean ratio of 2, the SCSE process with $\kappa=0$ could have a lag-1 autocorrelation between 0 and 0.707.  The INGARCH (1,1), on the other hand, with a variance to mean ratio of 2, must have an autocorrelation between 0.5 and 0.707.

\subsubsection{Spatial Covariance and Correlation}
The SCSE model also allows for limited spatial correlation Recalling that $\Sigma_{i,j}$ is the marginal covariance between $U(\boldsymbol{s_i},t)$ and $U(\boldsymbol{s_j},t)$, the spatial covariance between $Y(\boldsymbol{s_i},t)$ and $Y(\boldsymbol{s_j},t)$ is

\begin{align}
	& \mbox{Cov}\left[Y(\boldsymbol{s_i},t),Y(\boldsymbol{s_j},t):\forall i \neq j\right] =  \frac{\exp(2\alpha)}{1-(\eta+\kappa)^2}\left[\exp(\Sigma_{i,i}+\Sigma_{i,j}) -\exp(\Sigma_{i,i})\right] \label{eq:SpatCov}.
\end{align}
A proof of this is is given in Appendix B.  From \eqref{eq:SpatCov} it is clear that the spatial covariance is zero if the marginal covariance between $X(\boldsymbol{s_i})$ and $X(\boldsymbol{s_j})$ is zero.  However, if there is a non-zero marginal covariance between the spatial locations, $\alpha,\eta$ and $\kappa$ influence the spatial correlation in the data. As $\Sigma_{i,j}$ can be either positive or negative, the spatial covariance, unlike the temporal covariance, can be either positive or negative.

The spatial correlation is therefore
\begin{align}
	\mbox{Corr}(Y(\boldsymbol{s_i},t),Y(\boldsymbol{s_j},t)) & = \frac{\exp(2\alpha)\left(\exp(\Sigma_{i,i}+\Sigma_{i,j}) -\exp(\Sigma_{i,i})\right)}{\mbox{Var} (\exp(U(\boldsymbol{s_i},t)))+E[Y(\boldsymbol{s_i},t)]}\nonumber \\
 & = \frac{\left(\exp(\Sigma_{i,i}+\Sigma_{i,j}) -\exp(\Sigma_{i,i})\right)}{ \exp(2\Sigma_{i,i})-\exp(\Sigma_{i,i})+\exp(-\alpha+\frac{\Sigma_{i,i}}{2})\frac{1}{1-(\kappa+\eta)}}\label{eq:SpatCorr}.
\end{align}
The spatial correlation, as seen in \eqref{eq:SpatCorr}, only depends on $\eta$ and $\kappa$ through the expectation of $Y(\boldsymbol{s_i},t)$, however this is a potential limitation as this implies that the range of correlations depends on values of parameters other than the parameters of the CAR process.

\section{Bayesian Inference}\label{Sec:Bayes}
Bayesian analysis of the SCSE model can be accomplished through application of the techniques suggested by \cite{joseph}.  Letting the $\pi(\cdot)$ generically represent a density and $\pi(\cdot|\cdot)$ generically represent a conditional density, the joint prior distribution of the parameters in the model can be expressed as  $\pi(\boldsymbol{\theta})=\pi(\eta|\kappa)\pi(\kappa)\pi(\boldsymbol{\alpha})\pi(\sigma)\pi(\zeta)$ where we assume independence in our priors except for $\eta$ and $\kappa$ due to the restriction that $\eta+\kappa <1$.  Letting $U(\si,t) = Y(\si)+\epsilon(\si,t)$, the full conditional distribution of $\boldsymbol{\boldsymbol{\theta}}=\left(\eta,\kappa,\alpha,\zeta,\sigma^2\right)^T$ is

\begin{align}
	&\small\pi(\boldsymbol{\theta} | \boldsymbol{Y},\boldsymbol{U})\propto  \prod_{t=1}^T \left[\pi(\boldsymbol{Z}_t|\boldsymbol{\lambda}_t)\pi(\boldsymbol{\lambda}_t|\boldsymbol{\lambda}_{t-1},\boldsymbol{Y}_{t-1},\boldsymbol{\theta} ,\boldsymbol{U}_t)\pi(\boldsymbol{U}_t|\boldsymbol{\theta} )
	\pi(\boldsymbol{\lambda_0}|\boldsymbol{\theta})\pi(\boldsymbol{Y_0}|\boldsymbol{\lambda_0})\pi(\boldsymbol{\theta})\right]\label{eq:FullCondTheta},
\end{align}
and the full conditional of the complete vector $\boldsymbol{U}$ is

\begin{align}
	&\small\pi(\boldsymbol{U} | \boldsymbol{Y},\boldsymbol{\theta})\propto \prod_{t=1}^T\left[\pi(\boldsymbol{Y}_t|\boldsymbol{\lambda}_t)\pi(\boldsymbol{\lambda}_t|\boldsymbol{\lambda}_{t-1},\boldsymbol{Y}_{t-1},\boldsymbol{\theta} ,\boldsymbol{U}_t)\pi(\boldsymbol{U}_t|\boldsymbol{\theta} )\pi(\boldsymbol{\lambda_0}|\boldsymbol{\theta})\pi(\boldsymbol{Y_0}|\boldsymbol{\lambda_0})\right]
\end{align}
In order to do any form of Markov Chain Monte Carlo inference we must sample from the density of the full latent state, $\boldsymbol{U}$ which requires evaluations of 

\begin{align}
	\log(\boldsymbol{U}|\boldsymbol{\alpha},\sigma,\zeta) & \propto \frac{-T \times n}{2}\log(2 \pi) + \frac{1}{2} \log | \Sigma_f^{-1}(\boldsymbol{\theta})|\nonumber\\
	& - \frac{1}{2}(X-\alpha)^T\Sigma_f^{-1}(\boldsymbol{\theta})(X-\alpha) \label{eq:log Y1}.
\end{align}
Note that as our neighborhood structure is assumed to be constant  for all time periods so we can write $\Sigma_f(\boldsymbol{\theta})$ as the full space-time covariance matrix $\left(I_{T,T}\otimes (I_{{n},{n}}-\boldsymbol{C}\right)^{-1}\sigma^2+I_{n \times T} \sigma^2_{\epsilon}$.     The sparsity of the covariance structure means that the only computations of $\frac{1}{2}(X-\alpha)^T\Sigma_f^{-1}(\boldsymbol{\theta})(X-\alpha)$ that need to occur are for spatial neighbors. Therefore, the most difficult part of the computation of the log-density is the computation of the determinant, $\log | \Sigma_f^{-1}(\boldsymbol{\theta})|$.  However, the complicated notation of the covariance structure belies the fact that the precision matrix is both block diagonal and extremely sparse that greatly simplify computations of \eqref{eq:log Y1}.  The specific structure for $\Sigma^{-1}(\boldsymbol{\theta})=(\boldsymbol{I}-\boldsymbol{C})\frac{1}{\sigma^2}$ allows us to follow \cite{jin2005generalized}.  In particular, we have $\log|\Sigma^{-1}(\boldsymbol{\theta})|=\frac{n}{2\log\sigma^2}+\log|I_{n,n}-\zeta N|$ where $N$ is the neighborhood or adjacency matrix.  Letting $V \Lambda V^T$ be the spectral decomposition of $N$ we have $|I_{n,n}-\zeta N|=|V| |I_{n,n}-\zeta \Lambda| |V^T|=\prod_{j=1}^{n}\left(1-\zeta \chi_j\right)$ where $\chi_j$ are the eigenvalues of the neighborhood matrix.  Also, as $ \Sigma_f^{-1}(\boldsymbol{\theta})$ is block diagonal with each block being size $T \times T$ and having structure $\Sigma^{-1}$, it follows that $\log| \Sigma_f^{-1}(\boldsymbol{\theta})|=\frac{n \times T}{\log\sigma^2}+T\log| \Sigma^{-1}(\boldsymbol{\theta})|$

\begin{align}
	\log | \Sigma_f^{-1}(\boldsymbol{\theta})|&  = T \log | \Sigma^{-1}(\boldsymbol{\theta})|\\
	& \propto \frac{n \times T}{\log\sigma^2}+ T \sum_{j=1}^{n}(1-\zeta\chi_j) \label{eq:eig}
\end{align}

The greatest advantage of this approach is that the eigenvalues of the neighborhood matrix, $\boldsymbol{N}$ depend only on the neighborhood structure and do not depend on any parameters, therefore they can be computed ahead of time.  This means that we never need to deal with matrices of the size of  $\Sigma_f(\boldsymbol{\theta})$, rather we just need to find the eigenvalues for the neighborhood matrix.  This allows the model to be fit relatively quickly using a Gibbs algorithm combined with Hamiltonian Monte Carlo or similar methods currently implemented in the software package Stan \cite{carpenter2016stan}.

\subsection{Model Assessment and Simulation}\label{Sec:Simulations}

To conduct model assessment under the above framework we rely on posterior predictive P values, see e.g. \cite{gelman1996posterior}.  This technique samples new data sets after sampling parameters from the posterior distribution.  Statistics are calculated on the new data and compared to the statistics from the original dataset.  For each data set we calculate $T_1(\boldsymbol{Z})$, the spatial Moran's I statistic, $T_2(\boldsymbol{Z})$, the log of the variance to mean ratio, $T_3(\boldsymbol{Z})$, the variance of $Y(\boldsymbol{s_i},t)-Y(\boldsymbol{s_i},t-1)$ and $T_4$, the average sample lag-one Auto-regression. $T_1$ and $T_4$ capture the spatial and temporal structure in the data while $T_2$ and $T_3$ capture the variability and roughness at each location.  The roughness, or variance of the differences, captures how rare it is to see jumps in violence or crime. If this value is high, then we would expect the there to be large changes in crime some months but small change in crime, in the same location, between other months.  Note here we rely on the second order properties of the model for assessment as first order properties often are insufficient to identify spatio and spatio-temporal models which has led to the development of formal tests for second order properties \cite{diggle1991second}.

To demonstrate that the SCSE is able to differentiate between $\eta$ and $\kappa$ in the model and to demonstrated the use of the posterior predictive values in model assessment we simulated 50 realizations of \eqref{eq:timeseries2} from a $20 \times 20$ regular grid using a rook neighborhood structure allowing $\eta=0.2$ and $\kappa = 0.7$. We further set the spatial parameter $\zeta = 0.245$ and $\sigma = 0.5$. To decrease computation time we set $\sigma_{\epsilon} = 0$. For each realization we fit the model using the methodology outlined above and captured whether the true parameter was contained in the 95\% credible interval. We next simulated from the posterior, from each fitted model, and generated posterior predictive values for each simulation. In Table~\ref{Table:simPvalus} we provide the mean posterior predictive check from the 50 simulations for each statistic as well as the standard error demonstrating that, on average, the posterior predictive distribution is able to retain the four second order measures discussed above.

\begin{table}[!htp]
	\caption{Properties of Posterior Predictive Checks Simulating from SCSE} \label{Table:simPvalus} 
	\begin{center}
		\begin{tabular}{ |l|c|c|c| } 
			\hline
			 Posterior Predictive Check & Mean & Standard Error\\
			\hline 
			$p_1$ - Moran's I Statistic& 0.40 & 0.22  \\
			$p_2$ - Variance to Mean Ratio & 0.50 &0.21\\
			$p_3$ - Variance of $\Delta Y(\boldsymbol{s_i},t)$ & 0.50 &0.11 \\ 
			$p_4$ - AR(1) Value & 0.54 &0.09\\
			\hline
		\end{tabular}
	\end{center}
\end{table}

From the simulation, we found that 92\% of the time the posterior 95\% credible interval captured the generating parameters. Also, we see from Table~\ref{Table:simPvalus} we see that if the generating mechanism is a SCSE we would expect the posterior predictive checks for the second order properties to be centered around 0.50.

\section{Burglaries in Chicago} \label{sec:Chicago}
\begin{figure}[!htp]
	\includegraphics[ width=1.1\textwidth]{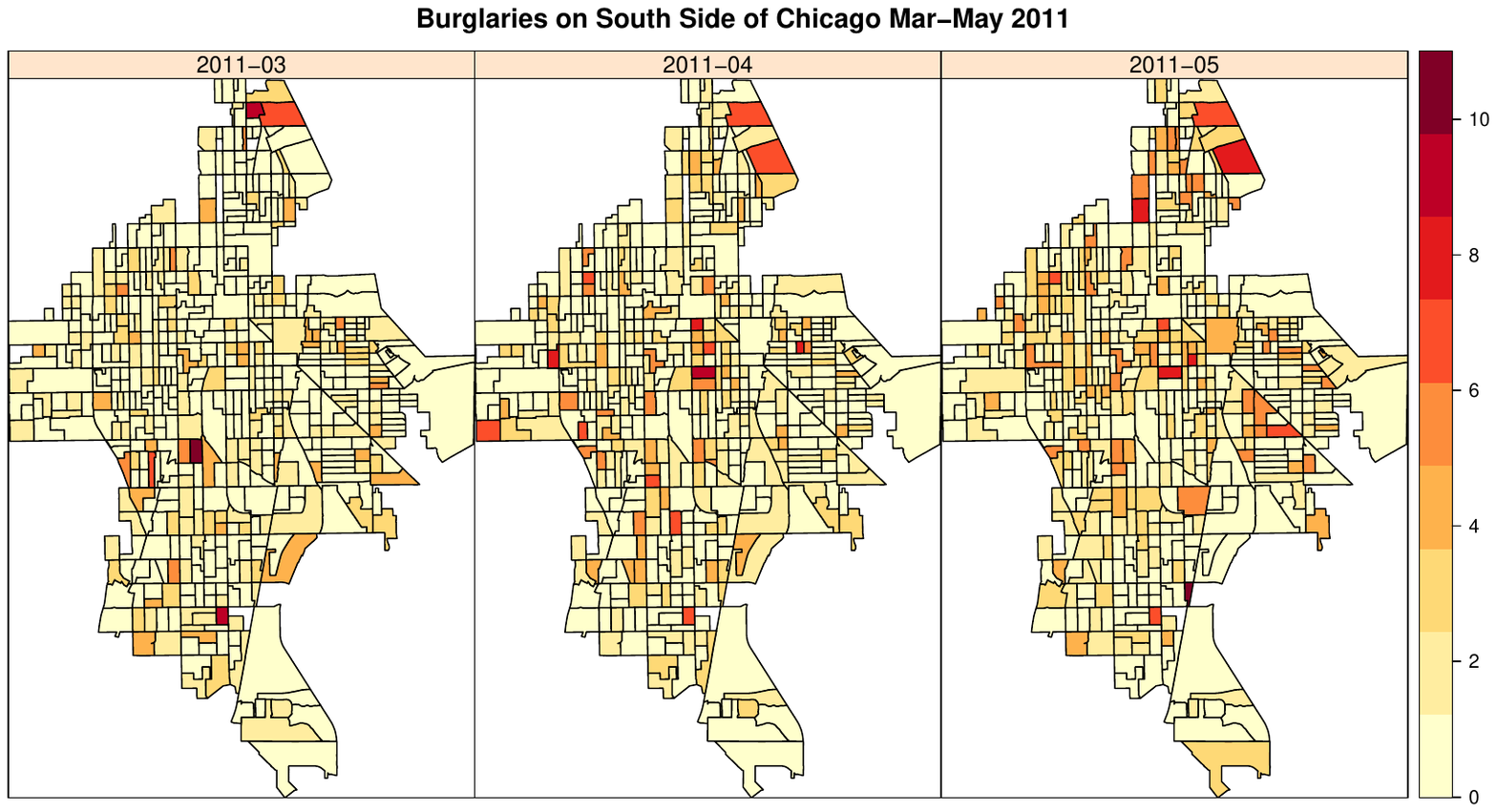}
	\caption{Three months of burglary counts within 552 Census block groups in South Chicago.  This area of Chicago is relatively racially and economically homogeneous}\label{fig:SouthSide}
\end{figure}

As a case study we consider a statistical model for burglaries in the south side of Chicago during 2010-2015 using crime data from the city of Chicago. As Chicago is one of the most racially and socio-economically segregated cities in America, we consider only the southside, a relatively racial and socio-economic homogeneous region depicted in figure \ref{fig:SouthSide}.  We aggregated the number of burglaries by Census block group and by month.  Within the south side of Chicago there are 552 census block groups resulting in a spatial domain of $\boldsymbol{s_i} \in \{\boldsymbol{s_1},\cdots,\boldsymbol{s_{552}}\}$ and temporal domain of $t \in \{1,2,\cdots,72\}$.  Two locations, $\boldsymbol{s_i}$ and $\boldsymbol{s_j}$, were considered neighbors when they shared a border.

While the geographic restriction to South Chicago eliminates some sources of socio-economic variability in the data, it does not eliminate all of it.  To account for this, we further consider covariates that address unique socio-economic and population characteristics for each region.  Specifically, we consider population, percentage of young males, per capita income, and percent unemployed. Unemployment, for example, has long been shown to have a relationship with crime, see e.g. \cite{britt1994crime} and \cite{raphael2001identifying}, the later showing property crime in particular has a strong relationship with unemployment.  

\begin{figure}[!htp]
\centering
	\includegraphics[width=100mm,scale=1]{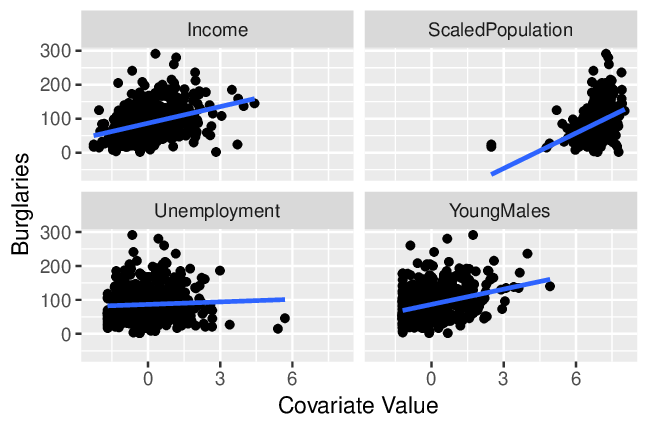}
	\caption{Exploratory, univariate, relationships between burglaries and income, unemployment,number of young males, and the log of the population. Here we see a weak positive relationship between the covariates and the response as evident in the trend lines.}\label{fig:heur}
\end{figure}

All potential covariates were obtained from the U.S. Census Tiger data available at \href{https://www.census.gov/geo/maps-data/data/tiger-data.html}{https://www.census.gov/geo/maps-data/data/tiger-data.html}.  The maximum number of burglaries in a month in a census block for this subset of the city is 17.  The variance to mean ratio in the data is 1.8, suggesting there is some overdispersion in the data.  There is both temporal and spatial clustering as evident by the average lag-one autocorrelation, .32, and the Moran's I statistic of .20. There is a clear seasonality trend in the data as well as a general downward trend from 2010-2015. This is not surprising and is consistent with previous findings in burglaries as seen in \cite{zhuang2019semiparametric}.  In order to account for this we preprocessed the data to remove the seasonality effect and the trend prior to estimating the impact of the spatial covariates and the process covariates.

\subsection{Self-Exciting Spatially Correlated Model for Chicago Burglaries}

In order to model this data we consider the model,
\begin{align}
& Y(\boldsymbol{s_i},t) \sim \mbox{Pois}(\lambda(\boldsymbol{s_i},t)) \label{eq:SCSECHI} \\
& E[Y(\boldsymbol{s_i},t)]=\lambda(\boldsymbol{s_i},t)\nonumber\\
& \boldsymbol{\lambda_t} = \exp(\boldsymbol{\epsilon_t}+\boldsymbol{X})+\eta \boldsymbol{Z_{t-1}}+\kappa \boldsymbol{\lambda_{t-1}}\nonumber\\
& \boldsymbol{X} \sim \mbox{Gau}\textit{} (\boldsymbol{\alpha},(\mathbf{N}-\zeta \mathbf{C})^{-1}\sigma_{sp}^2)\nonumber\\
& \boldsymbol{\epsilon}_t \sim \mbox{Gau} (\boldsymbol{0},(I_{n,n} \sigma_{\epsilon}^2)\nonumber.
\end{align}
If we again conceptualize $\lambda(s_i,t)$ as the tension at location $\boldsymbol{s_i}$ and month, $t$, this model says that the change in tension is due to five components.  The first is a baseline tension at location $s_i$ that can be explained through exogeneous covariates, $\alpha(s_i)$.  How $\alpha(s_i)$ manifests itself  is due to some small scale spatial structure captured in $\boldsymbol{T}$ as well as residual variability captured in $\boldsymbol{\epsilon}_t$.  The change in tension is also due to repeat victimization, $\eta$, as well as a natural decay over time captured in $\kappa$.  

To model the baseline tension at location $\boldsymbol{s_i}$ using available data from the US Census bureau, letting $\boldsymbol{\alpha}=(\alpha_{s_1},\cdots,\alpha_{s_{n}})^T$, we set

\begin{equation}
\alpha_{\boldsymbol{s_i}}=\exp\left(\beta_0+\beta_{pop} \log(\mbox{Pop}_{\boldsymbol{s_i}})+\beta_{ym}\mbox{Young Men}_{\boldsymbol{s_i}}+\beta_{wealth}\mbox{Wealth}_{\boldsymbol{s_i}}+\beta_{unemp}\mbox{Unemp}_{\boldsymbol{s_i}}\right) \label{eq:mean strucutre}.
\end{equation}

 As we are using a spatial structure based on Census block groups we must account for the fact that each location has a different number of spatial neighbors.  To adjust for this we use the weighted CAR model of \cite{besag1991bayesian} for $\boldsymbol{Y}$.  This spatial process assumes that the latent conditional variance for each location in the CAR model is $\frac{\sigma^2}{|N(s_i)|}$.  This process has a joint density given in the fourth line of \eqref{eq:SCSECHI} letting $\mathbf{N}$ be a diagonal matrix with entry $(1,1)$ equal to the number of neighbors of location $s_1$.  Recalling that $\mathbf{C}$ is the matrix that has entries 1 in position $(1,2)$ if $s_1$ and $s_2$ are geographically adjacent, the parameter space of $\zeta$ under this formulation is $\zeta \in (-1,1)$.   To account for the fact that along much of the parameter space of $\zeta$ the model is nearly unidentifiable we fix $\zeta$ near the edge of the parameter space ($\zeta=0.999$) (see \cite{wall2004close} for more issues on identifiability of the spatial parameter in a CAR model).

To complete the Bayesian inference we further need to place priors on all parameters in the model, except for $\zeta$ which we have fixed as above.  In order to minimize the impact of the prior selection on the posterior densities we select diffuse proper priors for $\beta_0,\beta_1,\beta_2,\eta,\kappa,\sigma^2$ and $\sigma_{\epsilon}^2$ and conducted sensitivity analysis to determine that the choice of prior had minimal impact on the results. Diffuse proper priors are vague priors that seek to limit the impact of prior selection on the posterior. However, unlike improper priors, they are given valid distributions ensuring the posterior distribution remains valid. More discussion on the choice of priors is given in Section~\ref{sec:FitCompare}.

\subsection{Alternate Models for Chicago Burglaries}

An alternate model that describes the spread of burglaries is similar to the model of \cite{short2008statistical}.  We might assume that the change in the rate of burglaries at a location is a function of a base attractiveness due to unique geographical features at that location, $\alpha_{\boldsymbol{s_i}}$ as described in \eqref{eq:mean strucutre}, a natural decay over time, $\chi$, and repeat victimization, $\eta$.  In that model, the authors further considered a spatial spread, parameterized by $\psi$ and motivated through a reaction-diffusion difference equation.  In \eqref{eq:DiscreteModel} the use of $\psi$ was motivated by the belief that regions of high violence will spread the other regions. This is similar to a heat equation where high temperatures spread to cooler areas. These assumptions lead to the (stochastic) difference equation

\begin{equation}
	\lambda(\boldsymbol{s_i},t+1)-\lambda(\boldsymbol{s_i},t)=\alpha_{\boldsymbol{s_i}}-\chi \lambda(\boldsymbol{s_i},t)+\frac{\psi}{|N_{s_i}|}\sum_{s_j \in N_{i}} \left[\lambda(\boldsymbol{s_j},t- 1)-\lambda(\boldsymbol{s_i},t- 1)\right]+\eta Y(\boldsymbol{s_i},t)\label{eq:DiscreteModel}.
\end{equation}
In practice, though when applied to the Chicago dataset, $\psi$ was found to be zero, potentially due to the choice in how we aggregated time and space.  If we fix $\psi=0$ \eqref{eq:DiscreteModel} is now an INGARCH model with $\kappa=(1-\chi)$.

Finally, we compare the model to what is sometimes called the CAR ANOVA model, e.g. \cite{lee2018spatio}.  Here we let,

\begin{align}
& Y(\boldsymbol{s_i},t) \sim \mbox{Pois}(\lambda(\boldsymbol{s_i},t))\nonumber  \\
& E[Y(\boldsymbol{s_i},t)]=\lambda(\boldsymbol{s_i},t)\nonumber\\
& \log\left(\lambda_{\si,t}\right) = \boldsymbol{\alpha}+\phi_{\si}+\delta_t\nonumber\\
& \phi_{\si}| \phi_{-\si} \sim N\left(\frac{\sum_{\boldsymbol{s_j}\in N(\si)}\phi_{s_j}}{|N(\si)|}, \frac{\tau^2_s}{|N(\si)|}\right)\nonumber\\
& \delta_t| \delta_{-t} \sim N\left(\rho \delta_{t-1},\frac{\tau^2_t}{\rho} \right)
\label{eq:CAR ANOVA}.
\end{align}

To complete the specification, This model assumes that the log intensity can be decomposed into a spatially varying random effect and a temporally varying auto-regressive component.  Large scale spatial structure can still be accounted for in this model by placing structure on $\boldsymbol{\alpha}$.  Although not presented in this manuscript we also considered statistical models that allowed the spatial field to vary temporally similar to the CAR AR given in \cite{lee2018spatio}, these models did not significantly improve over the CAR ANOVA model presented above.  Model \eqref{eq:CAR ANOVA} was fit using the CARBayesST package in R, \cite{lee2015carbayesst}.

\subsection{Model Fit and Comparison}\label{sec:FitCompare}

To the extent possible we used vague proper priors for all models. For instance in both \eqref{eq:SCSECHI} and \eqref{eq:DiscreteModel} we place independent normal priors with mean zero and standard deviation 10 on each of the $\beta$ terms and place $\beta(0.5,0.5)$ on each of the $\eta$ and $\kappa$ parameters. In \eqref{eq:SCSECHI} both of the $\sigma_{\epsilon}$ and $\sigma_{sp}$ parameters were given a half-Cauchy prior. The same principles were used in fitting \eqref{eq:CAR ANOVA}.

95\% credible intervals found from fitting \eqref{eq:SCSECHI} using the procedure outlined above and a similar Bayesian inference for \eqref{eq:DiscreteModel} are given in Table \ref{Table:ResultsChi}.   $\hat{R}$ and visual examination of the chains indicated no evidence that they had not converged. Divergence transitions for the Markov chains were also checked and eliminated. The CAR ANOVA model was fit using the R package CARBayesST and the Geweke Diagnostic was monitored for convergence.

\begin{table}[!htp]
	\begin{center}
		\begin{tabular}{ |c|c|c|c| } 
			\hline
			Parameter &SCSE Model & CAR ANOVA & INGARCH(1,1) \\
			\hline 
			$\beta_0$ & (-3.3,-1.0)& (-2.3,-0.8) & (-4.2,-3.4) \\
			$\beta_{pop}$ & (0.11,0.34) & (0.05,0.30)&(0.33,0.46)\\
			$\beta_{unemp}$ & (-0.75,0.17)& (-0.8,0.15)&(0.06,0.09)\\
			$\beta_{wealth}$&(0.05,0.16) & (0.06,0.18)& (-0.04,0.01)\\
			$\beta_{males}$ & (0.006,0.07)& (0.005,0.07) & (0.002,0.03) \\
			$\eta$ & (0.04,0.07) & - & (0.22,0.24)\\
			$\kappa$ & (0.31,0.39)& - & (0.44,0.48)\\
			$ \rho$ & - & (0.60, 0.94) & - \\
			$\sigma_{sp}^2$ & (0.40,0.54)& - & - \\
			$\sigma_{\epsilon}^2$& (0.40,0.47)& - & - \\
			$\tau_s^2$& - & (0.47,0.62) & - \\
			$ \tau^2_t$ & - &(0.01,0.03) & - \\
			\hline
		\end{tabular}
	\end{center}
	\caption{95\% Credible Intervals for parameters of the SCSE model given in \eqref{eq:SCSECHI} and INGARCH(1,1) models applied to the Chicago burglary data.}\label{Table:ResultsChi}
\end{table}

The SCSE parameters suggest that the $\alpha_{\boldsymbol{s_i}}$ process considered manifests itself differently at each unique spatio-temporal location even when residual spatial variation is accounted for in the CAR model.  In other words, there may exist small-scale spatial effects that are captured in the CAR model as well as unique characteristics of each location that the CAR model does not fully explain.

In examining the fit of the models we first visually examined the posterior mean of $\lambda_{\si,t}$ from all three models and compared them to the observed $Y_{\si,t}$ values.

\begin{figure}[!htp]
\centering
	\includegraphics[ width=0.7\textwidth]{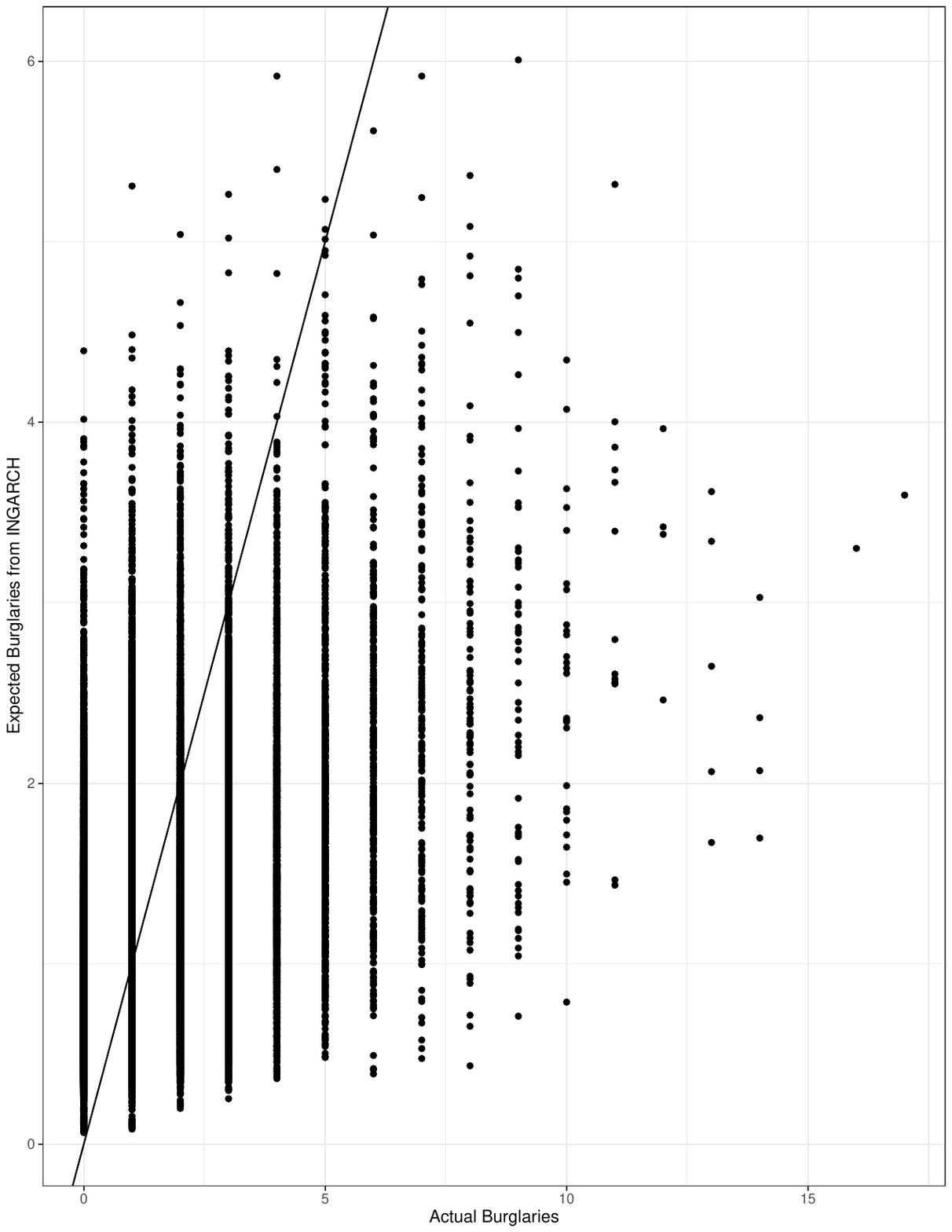}
	\caption{Posterior mean from $\lambda_{\si,t}$ found from the INGARCH model plotted against $Y_{\si,t}$}\label{fig:INGARCHFit}
\end{figure}

\begin{figure}[!htp]
	\includegraphics[ width=0.7\textwidth]{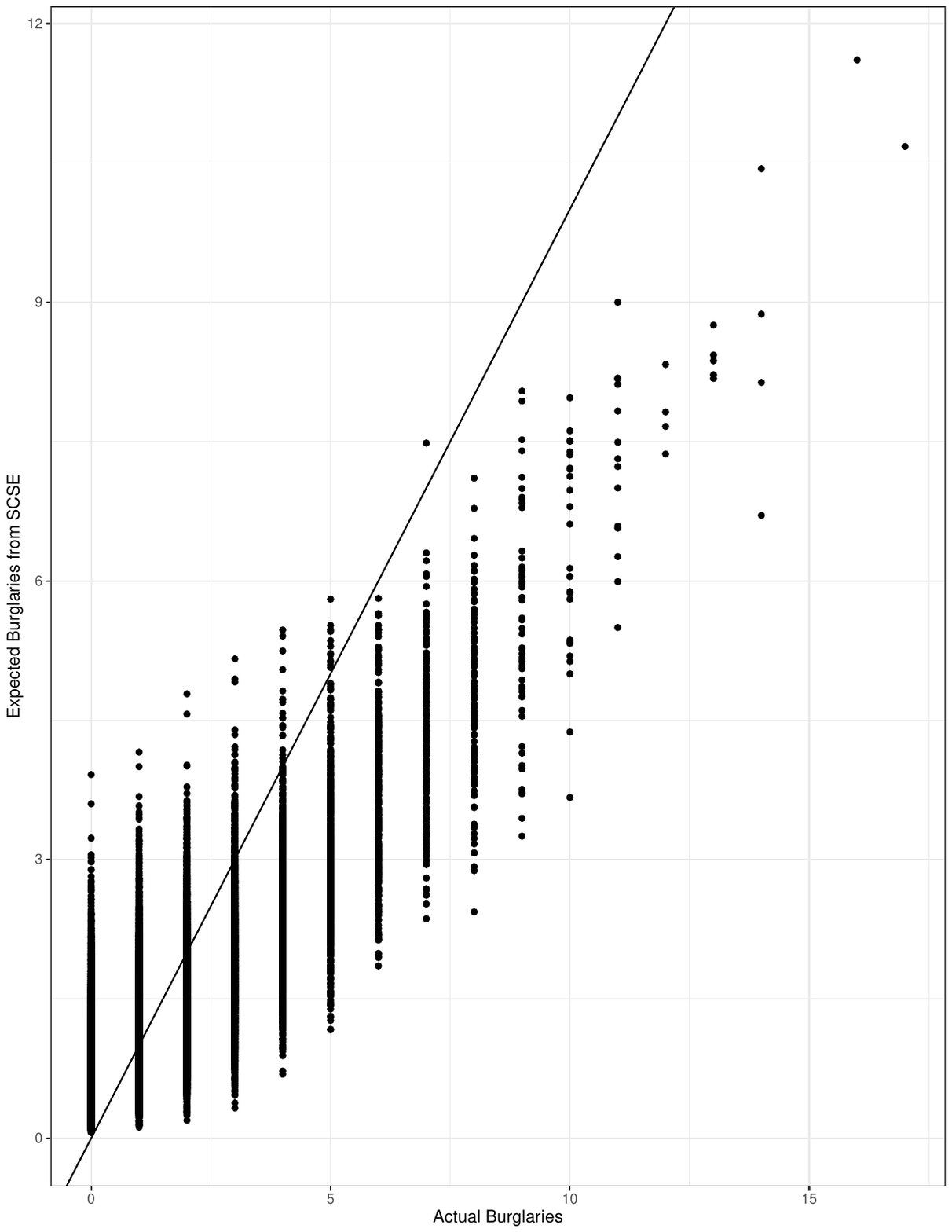}
	\caption{Posterior mean from $\lambda_{\si,t}$ found from the SCSE model plotted against $Y_{\si,t}$}\label{fig:SCSEFit}
\end{figure}

\begin{figure}[!htp]
	\includegraphics[ width=0.7\textwidth]{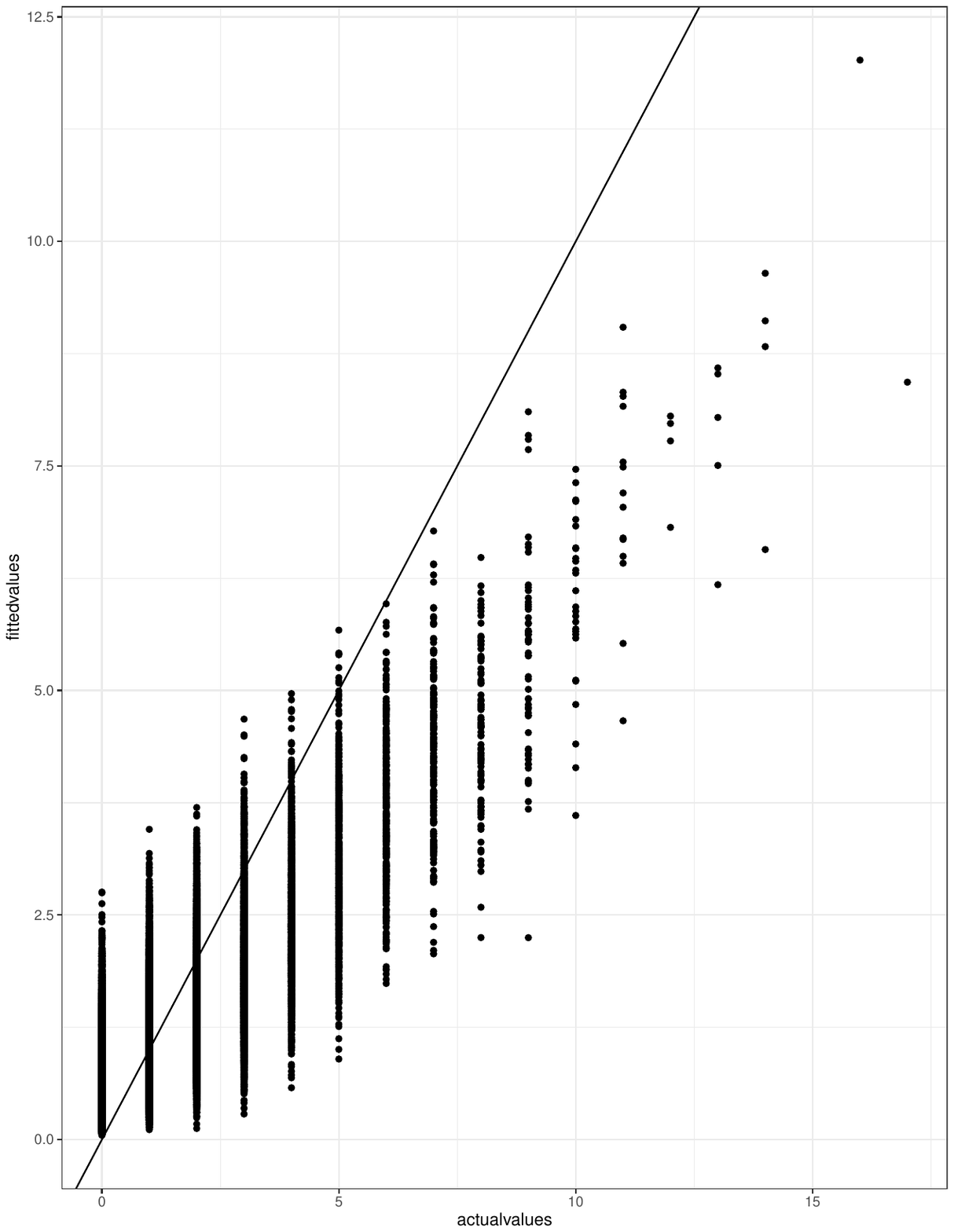}
	\caption{Posterior mean from $\lambda_{\si,t}$ found from the CAR ANOVA model plotted against $Y_{\si,t}$}\label{fig:CARAnovaFit}
\end{figure}

From Figures~\ref{fig:INGARCHFit},\ref{fig:SCSEFit}, and \ref{fig:CARAnovaFit} we see that there appears to be little difference between the SCSE model and the CAR ANOVA model however both models would appear to be preferable to the INGARCH model. In comparing the SCSE model to the CAR ANOVA model, 52\% of the time the CAR ANOVA model yielded a posterior mean that was closer to the observed value; again suggesting there is very little difference between the two models in examining the first order properties.

To further differentiate the models, though, we can also look at the second order properties. As discussed above in Section~\ref{Sec:Simulations}, we would expect the posterior predictive values for the second order properties to yield results close to 0.50.

\begin{table}[!htp]
	\caption{Posterior Predictive Checks for Chicago Burglary Data} \label{Table:Pvals} 
	\begin{center}
		\begin{tabular}{ |l|c|c|c| } 
			\hline
			&SCSE Model & CAR ANOVA Model & INGARCH(1,1)\\
			\hline 
			$p_1$ - Moran's I Statistic& 0.52 & 0.004 & 0 \\
			$p_2$ - Variance to Mean Ratio & 0.67 &0 & 0\\
			$p_3$ - Variance of $\Delta Y(\boldsymbol{s_i},t)$ & 0.60 &0 & 0.10 \\ 
			$p_4$ - AR(1) Value & 0.56 &0 & 0.23\\
			\hline
		\end{tabular}
	\end{center}
\end{table}
As seen in the Table~\ref{Table:Pvals}, not surprisingly the INGARCH process given in \eqref{eq:DiscreteModel} is not able to capture the spatial structure nor the variance to mean ratio in the data. However we also see here that the CAR ANOVA model fails to replicate the second order properties as well. The use of a model where the spatio-temporal structure is entirely placed on a latent state, the CAR ANOVA model, underestimates the spatial structure, the temporal structure, the variance to mean ratio as well as a measure of how ``choppy" the data are.  These models tend to over smooth the data failing to retain the unique characteristics that are often present in crime or violence models. Some over smoothing may be avoided through adaptive neighborhood structures as in \cite{rodrigues2012bayesian} albeit at a cost in computational complexity. Such an approach could likely be extended to account for both spatial as well as temporal structure though we do not further pursue this further in this manuscript.

As both the INGARCH model and the SCSE model have interpretations to may lead to policy decisionts as discussed below, it's worth further comparing these to models in particular. To see how well the SCSE model given in \eqref{eq:SCSECHI} replicates other characteristics of the data we also conducted posterior predictive checks of the maximum observed value (p=0.78 vs p=0 for INGARCH) and the number of zeros in the data set (p=0.40 vs p=0.20 for INGARCH).  Finally, we note that if the only posterior predictive check that was performed was on the AR(1) statistic, both the SCSE and the INGARCH models perform adequately (p=0.56 and p=0.23). Overall, while it is plausible that a model akin to \eqref{eq:SCSECHI} was the true generating mechanism, it is highly unlikely that the INGARCH model given in \eqref{eq:mean strucutre} generated the data.  

\section{Discussion}

The potential largest implication of the above analysis is that if the INGARCH model was fit, one would be tempted to conclude that there exists significant self-excitement due to the high $\eta$ value which is not present in the SCSE formulation.  This may lead to the potentially erroneous conclusion that there is repeat victimization present in the data.   However, since the SCSE model appears to fit the data better, the apparent self-excitement may actually be misspecification of the error structure in the model.  As the self-excitement parameter captures either repeat burglaries or burglaries motivated by a previous successful action, concluding the existence of self-excitement may have policy implications if the model was used in practice.  Contrary to this, previous research in \cite{polvi1991time} suggested that a burglarized home had an elevated risk of another burglary within six weeks, however the elevated risk, may in fact, be explained through unexplained spatial correlation as demonstrated in our analysis.  Policy implications of concluding repeat victimization, or self-excitement, is present in an area are discussed in \cite{pease1998repeat}.  In \cite{reinhart2018self}, the authors found that omitting a spatial covariate impacted conclusions on self-excitation.  In our work, we found more generally incorrectly specifying the structure of the background rate also biases findings regarding self-excitation.  As pointed out in \cite{reinhart2018self}, if the basis of our inference is on the covariates, the inclusion of the spatially correlated $X(\si)$ or $U(\si)$ term may bias covariate terms.

While the SCSE has convenient properties, the largest limitation is the computational time required to fit the model. In general, even using MCMC techniques such as Hamiltonian monte carlo as in \cite{gelman2015stan}, even modest sized spatial fields are slow to converge. While this can be aided through the use of high performance computers, practitioners who are conducting exploratory model building may find these unfeasible. Further, the SCSE is difficult to use with temporally varying covariates. As demonstrated in \cite{zhuang2019semiparametric}, incorporating seasonality is critical in understanding the temporal trend of burglaries. While we treat this as a fixed term in our model, extensions allowing for temporal varying covariates are necessary for a complete treatment of crime. Overall, though, it is important to realize that self-excitation, in a statistical model, may be a misspecification of error and it is insufficient to only examine the first order properties of the fitted model. 

\section{Conclusion}
 
In this manuscript we formulated a statistical model that contains both latent structure spatial dependence and data model temporal dependence extending earlier work of \cite{ferland2006integer}, \cite{fokianos2009poisson}, and \cite{davis2016theory}.  We also demonstrated how such models can arise from stochastic difference equations where the number of new arrivals into a process are no longer static but rather are themselves stochastic and how these assumptions are consistent with beliefs on how violence and crimes evolve over space and time. The resulting SCSE model is novel in its combination of both data model and latent model dependence and greatly extends the uses of the INGARCH(1,1) process. 

While not intending to be a complete treatment of Burglary in Chicago, the above demonstration does show that the SCSE process outperforms classical spatio-temporal models while offering interpretability and accounting for the possibility of data model dependence after conditioning on a latent process.  R code for fitting both the INGARCH model and the SCSE model given in \eqref{eq:SCSECHI} is available at \url{https://github.com/nick3703/Chicago-Data}.  

Extensions of this methodology should focus on decreasing computational time through techniques such as those used in Integrated Nested Laplace Approximation \cite{rue2009approximate}. As seen in \cite{python2019bayesian} this technique can be used to fit massive datasets likely greatly decreasing computational time. Further, an in depth study of how data clusters in space and time would be greatly beneficial. It is clear that both latent Gaussian models as well as self-exciting models are able to generate spatial as well as temporal structure, but as the interpretation of the models may be grossly different, it is important to understand why and how the data manifest differently in the two cases.

Lastly, as evident in the example, the restriction to a spatial CAR process for the latent variable $X(\si)$ is not necessary in practice.  Oftentimes a CAR specification is preferable as it is easier to model real-world phenomena conditionally.  On the other hand, in \cite{clark2018modeling}, both a Simultaneous Auto-Regressive (SAR) and a Vector Auto-Regressive (VAR) specification were used to model the latent variable.  In the former case, the model properties derived above will still hold, however in the later case the latent state also contained temporal covariance.  It is not immediately obvious that the ergodic properties of the model still exist if the latent process is allowed to propagate over time in this manner making derivations of the second order process, then, more difficult.

\section*{Acknowledgement}
We would like to extend our gratitude to three anonymous reviewers and an associate editor whose comments and feedback greatly improved this manuscript.

\section*{Declaration of Interests}
None

\bibliographystyle{elsarticle-harv}
\bibliography{sample.bib}

\begin{thebibliography}{48}
\expandafter\ifx\csname natexlab\endcsname\relax\def\natexlab#1{#1}\fi
\providecommand{\url}[1]{\texttt{#1}}
\providecommand{\href}[2]{#2}
\providecommand{\path}[1]{#1}
\providecommand{\DOIprefix}{doi:}
\providecommand{\ArXivprefix}{arXiv:}
\providecommand{\URLprefix}{URL: }
\providecommand{\Pubmedprefix}{pmid:}
\providecommand{\doi}[1]{\href{http://dx.doi.org/#1}{\path{#1}}}
\providecommand{\Pubmed}[1]{\href{pmid:#1}{\path{#1}}}
\providecommand{\bibinfo}[2]{#2}
\ifx\xfnm\relax \def\xfnm[#1]{\unskip,\space#1}\fi
\bibitem[{Aitchison and Ho(1989)}]{aitchison1989multivariate}
\bibinfo{author}{Aitchison, J.}, \bibinfo{author}{Ho, C.},
  \bibinfo{year}{1989}.
\newblock \bibinfo{title}{The multivariate poisson-log normal distribution}.
\newblock \bibinfo{journal}{Biometrika} \bibinfo{volume}{76},
  \bibinfo{pages}{643--653}.
\bibitem[{Athreya and Pantula(1986)}]{athreya1986mixing}
\bibinfo{author}{Athreya, K.B.}, \bibinfo{author}{Pantula, S.G.},
  \bibinfo{year}{1986}.
\newblock \bibinfo{title}{Mixing properties of harris chains and autoregressive
  processes}.
\newblock \bibinfo{journal}{Journal of applied probability}
  \bibinfo{volume}{23}, \bibinfo{pages}{880--892}.
\bibitem[{Augustin et~al.(2006)Augustin, McNicol and
  Marriott}]{augustin2006using}
\bibinfo{author}{Augustin, N.H.}, \bibinfo{author}{McNicol, J.},
  \bibinfo{author}{Marriott, C.A.}, \bibinfo{year}{2006}.
\newblock \bibinfo{title}{Using the truncated auto-poisson model for spatially
  correlated counts of vegetation}.
\newblock \bibinfo{journal}{Journal of agricultural, biological, and
  environmental statistics} \bibinfo{volume}{11}, \bibinfo{pages}{1--23}.
\bibitem[{Bernasco and Block(2011)}]{bernasco2011robberies}
\bibinfo{author}{Bernasco, W.}, \bibinfo{author}{Block, R.},
  \bibinfo{year}{2011}.
\newblock \bibinfo{title}{Robberies in chicago: A block-level analysis of the
  influence of crime generators, crime attractors, and offender anchor points}.
\newblock \bibinfo{journal}{Journal of Research in Crime and Delinquency}
  \bibinfo{volume}{48}, \bibinfo{pages}{33--57}.
\bibitem[{Besag(1974)}]{besag1974spatial}
\bibinfo{author}{Besag, J.}, \bibinfo{year}{1974}.
\newblock \bibinfo{title}{Spatial interaction and the statistical analysis of
  lattice systems}.
\newblock \bibinfo{journal}{Journal of the Royal Statistical Society. Series B
  (Methodological)} , \bibinfo{pages}{192--236}.
\bibitem[{Besag et~al.(1991)Besag, York and Molli{\'e}}]{besag1991bayesian}
\bibinfo{author}{Besag, J.}, \bibinfo{author}{York, J.},
  \bibinfo{author}{Molli{\'e}, A.}, \bibinfo{year}{1991}.
\newblock \bibinfo{title}{Bayesian image restoration, with two applications in
  spatial statistics}.
\newblock \bibinfo{journal}{Annals of the institute of statistical mathematics}
  \bibinfo{volume}{43}, \bibinfo{pages}{1--20}.
\bibitem[{Brantingham et~al.(2009)Brantingham, Brantingham, Vajihollahi and
  Wuschke}]{brantingham2009crime}
\bibinfo{author}{Brantingham, P.L.}, \bibinfo{author}{Brantingham, P.J.},
  \bibinfo{author}{Vajihollahi, M.}, \bibinfo{author}{Wuschke, K.},
  \bibinfo{year}{2009}.
\newblock \bibinfo{title}{Crime analysis at multiple scales of aggregation: a
  topological approach}, in: \bibinfo{booktitle}{Putting crime in its place}.
  \bibinfo{publisher}{Springer}, pp. \bibinfo{pages}{87--107}.
\bibitem[{Britt(1994)}]{britt1994crime}
\bibinfo{author}{Britt, C.L.}, \bibinfo{year}{1994}.
\newblock \bibinfo{title}{Crime and unemployment among youths in the united
  states, 1958-1990}.
\newblock \bibinfo{journal}{American Journal of Economics and Sociology}
  \bibinfo{volume}{53}, \bibinfo{pages}{99--109}.
\bibitem[{Carpenter et~al.(2016)Carpenter, Gelman, Hoffman, Lee, Goodrich,
  Betancourt, Brubaker, Guo, Li and Riddell}]{carpenter2016stan}
\bibinfo{author}{Carpenter, B.}, \bibinfo{author}{Gelman, A.},
  \bibinfo{author}{Hoffman, M.}, \bibinfo{author}{Lee, D.},
  \bibinfo{author}{Goodrich, B.}, \bibinfo{author}{Betancourt, M.},
  \bibinfo{author}{Brubaker, M.A.}, \bibinfo{author}{Guo, J.},
  \bibinfo{author}{Li, P.}, \bibinfo{author}{Riddell, A.},
  \bibinfo{year}{2016}.
\newblock \bibinfo{title}{Stan: A probabilistic programming language}.
\newblock \bibinfo{journal}{Journal of Statistical Software}
  \bibinfo{volume}{20}, \bibinfo{pages}{1--37}.
\bibitem[{Clark et~al.(2018)Clark, Dixon et~al.}]{clark2018modeling}
\bibinfo{author}{Clark, N.J.}, \bibinfo{author}{Dixon, P.M.}, et~al.,
  \bibinfo{year}{2018}.
\newblock \bibinfo{title}{Modeling and estimation for self-exciting
  spatio-temporal models of terrorist activity}.
\newblock \bibinfo{journal}{The Annals of Applied Statistics}
  \bibinfo{volume}{12}, \bibinfo{pages}{633--653}.
\bibitem[{Cressie and Wikle(2015)}]{cressie2015statistics}
\bibinfo{author}{Cressie, N.}, \bibinfo{author}{Wikle, C.K.},
  \bibinfo{year}{2015}.
\newblock \bibinfo{title}{Statistics for spatio-temporal data}.
\newblock \bibinfo{publisher}{John Wiley \& Sons}.
\bibitem[{Davis and Liu(2016)}]{davis2016theory}
\bibinfo{author}{Davis, R.A.}, \bibinfo{author}{Liu, H.}, \bibinfo{year}{2016}.
\newblock \bibinfo{title}{Theory and inference for a class of nonlinear models
  with application to time series of counts}.
\newblock \bibinfo{journal}{Statistica Sinica} \bibinfo{volume}{26},
  \bibinfo{pages}{1673--1707}.
\bibitem[{Diggle and Chetwynd(1991)}]{diggle1991second}
\bibinfo{author}{Diggle, P.J.}, \bibinfo{author}{Chetwynd, A.G.},
  \bibinfo{year}{1991}.
\newblock \bibinfo{title}{Second-order analysis of spatial clustering for
  inhomogeneous populations}.
\newblock \bibinfo{journal}{Biometrics} , \bibinfo{pages}{1155--1163}.
\bibitem[{Fefferman et~al.(2005)Fefferman, O'Neil and
  Naumova}]{fefferman2005confidentiality}
\bibinfo{author}{Fefferman, N.H.}, \bibinfo{author}{O'Neil, E.A.},
  \bibinfo{author}{Naumova, E.N.}, \bibinfo{year}{2005}.
\newblock \bibinfo{title}{Confidentiality and confidence: is data aggregation a
  means to achieve both?}
\newblock \bibinfo{journal}{Journal of public health policy}
  \bibinfo{volume}{26}, \bibinfo{pages}{430--449}.
\bibitem[{Ferland et~al.(2006)Ferland, Latour and Oraichi}]{ferland2006integer}
\bibinfo{author}{Ferland, R.}, \bibinfo{author}{Latour, A.},
  \bibinfo{author}{Oraichi, D.}, \bibinfo{year}{2006}.
\newblock \bibinfo{title}{Integer-valued garch process}.
\newblock \bibinfo{journal}{Journal of Time Series Analysis}
  \bibinfo{volume}{27}, \bibinfo{pages}{923--942}.
\bibitem[{Fokianos et~al.(2009)Fokianos, Rahbek and
  Tj{\o}stheim}]{fokianos2009poisson}
\bibinfo{author}{Fokianos, K.}, \bibinfo{author}{Rahbek, A.},
  \bibinfo{author}{Tj{\o}stheim, D.}, \bibinfo{year}{2009}.
\newblock \bibinfo{title}{Poisson autoregression}.
\newblock \bibinfo{journal}{Journal of the American Statistical Association}
  \bibinfo{volume}{104}, \bibinfo{pages}{1430--1439}.
\bibitem[{Gelman et~al.(2015)Gelman, Lee and Guo}]{gelman2015stan}
\bibinfo{author}{Gelman, A.}, \bibinfo{author}{Lee, D.}, \bibinfo{author}{Guo,
  J.}, \bibinfo{year}{2015}.
\newblock \bibinfo{title}{Stan: A probabilistic programming language for
  bayesian inference and optimization}.
\newblock \bibinfo{journal}{Journal of Educational and Behavioral Statistics}
  \bibinfo{volume}{40}, \bibinfo{pages}{530--543}.
\bibitem[{Gelman et~al.(1996)Gelman, Meng and Stern}]{gelman1996posterior}
\bibinfo{author}{Gelman, A.}, \bibinfo{author}{Meng, X.L.},
  \bibinfo{author}{Stern, H.}, \bibinfo{year}{1996}.
\newblock \bibinfo{title}{Posterior predictive assessment of model fitness via
  realized discrepancies}.
\newblock \bibinfo{journal}{Statistica sinica} , \bibinfo{pages}{733--760}.
\bibitem[{Genest and Ne{\v{s}}lehov{\'a}(2007)}]{genest2007primer}
\bibinfo{author}{Genest, C.}, \bibinfo{author}{Ne{\v{s}}lehov{\'a}, J.},
  \bibinfo{year}{2007}.
\newblock \bibinfo{title}{A primer on copulas for count data}.
\newblock \bibinfo{journal}{ASTIN Bulletin: The Journal of the IAA}
  \bibinfo{volume}{37}, \bibinfo{pages}{475--515}.
\bibitem[{Goicoa et~al.(2016)Goicoa, Ugarte, Etxeberria and
  Militino}]{goicoa2016age}
\bibinfo{author}{Goicoa, T.}, \bibinfo{author}{Ugarte, M.},
  \bibinfo{author}{Etxeberria, J.}, \bibinfo{author}{Militino, A.},
  \bibinfo{year}{2016}.
\newblock \bibinfo{title}{Age--space--time car models in bayesian disease
  mapping}.
\newblock \bibinfo{journal}{Statistics in medicine} .
\bibitem[{Hawkes(1971)}]{hawkes1971spectra}
\bibinfo{author}{Hawkes, A.G.}, \bibinfo{year}{1971}.
\newblock \bibinfo{title}{Spectra of some self-exciting and mutually exciting
  point processes}.
\newblock \bibinfo{journal}{Biometrika} , \bibinfo{pages}{83--90}.
\bibitem[{Heinen(2003)}]{heinen2003modelling}
\bibinfo{author}{Heinen, A.}, \bibinfo{year}{2003}.
\newblock \bibinfo{title}{Modelling time series count data: an autoregressive
  conditional poisson model} .
\bibitem[{Heinen and Rengifo(2007)}]{heinen2007multivariate}
\bibinfo{author}{Heinen, A.}, \bibinfo{author}{Rengifo, E.},
  \bibinfo{year}{2007}.
\newblock \bibinfo{title}{Multivariate autoregressive modeling of time series
  count data using copulas}.
\newblock \bibinfo{journal}{Journal of Empirical Finance} \bibinfo{volume}{14},
  \bibinfo{pages}{564--583}.
\bibitem[{Hu et~al.(2018)Hu, Zhu, Duan and Guo}]{hu2018urban}
\bibinfo{author}{Hu, T.}, \bibinfo{author}{Zhu, X.}, \bibinfo{author}{Duan,
  L.}, \bibinfo{author}{Guo, W.}, \bibinfo{year}{2018}.
\newblock \bibinfo{title}{Urban crime prediction based on spatio-temporal
  bayesian model}.
\newblock \bibinfo{journal}{PloS one} \bibinfo{volume}{13},
  \bibinfo{pages}{e0206215}.
\bibitem[{Jin et~al.(2005)Jin, Carlin and Banerjee}]{jin2005generalized}
\bibinfo{author}{Jin, X.}, \bibinfo{author}{Carlin, B.P.},
  \bibinfo{author}{Banerjee, S.}, \bibinfo{year}{2005}.
\newblock \bibinfo{title}{Generalized hierarchical multivariate car models for
  areal data}.
\newblock \bibinfo{journal}{Biometrics} \bibinfo{volume}{61},
  \bibinfo{pages}{950--961}.
\bibitem[{Joseph(2016)}]{joseph}
\bibinfo{author}{Joseph, M.}, \bibinfo{year}{2016}.
\newblock \bibinfo{title}{Exact sparse car models in stan}.
\newblock
  \bibinfo{howpublished}{\url{http://mc-stan.org/users/documentation/case-studies/mbjoseph-CARStan.html}}.
\newblock \bibinfo{note}{Accessed: 2017-07-17}.
\bibitem[{Kaiser and Cressie(1997)}]{kaiser1997modeling}
\bibinfo{author}{Kaiser, M.S.}, \bibinfo{author}{Cressie, N.},
  \bibinfo{year}{1997}.
\newblock \bibinfo{title}{Modeling poisson variables with positive spatial
  dependence}.
\newblock \bibinfo{journal}{Statistics \& Probability Letters}
  \bibinfo{volume}{35}, \bibinfo{pages}{423--432}.
\bibitem[{Laub et~al.(2015)Laub, Taimre and Pollett}]{laub2015hawkes}
\bibinfo{author}{Laub, P.J.}, \bibinfo{author}{Taimre, T.},
  \bibinfo{author}{Pollett, P.K.}, \bibinfo{year}{2015}.
\newblock \bibinfo{title}{Hawkes processes}.
\newblock \bibinfo{journal}{arXiv preprint arXiv:1507.02822} .
\bibitem[{Lee et~al.(2015)Lee, Rushworth and Napier}]{lee2015carbayesst}
\bibinfo{author}{Lee, D.}, \bibinfo{author}{Rushworth, A.},
  \bibinfo{author}{Napier, G.}, \bibinfo{year}{2015}.
\newblock \bibinfo{title}{Carbayesst: spatio-temporal generalised linear mixed
  models for areal unit data}.
\newblock \bibinfo{journal}{R package version} \bibinfo{volume}{2}.
\bibitem[{Lee et~al.(2018)Lee, Rushworth and Napier}]{lee2018spatio}
\bibinfo{author}{Lee, D.}, \bibinfo{author}{Rushworth, A.},
  \bibinfo{author}{Napier, G.}, \bibinfo{year}{2018}.
\newblock \bibinfo{title}{Spatio-temporal areal unit modelling in r with
  conditional autoregressive priors using the carbayesst package}.
\newblock \bibinfo{journal}{Journal of Statistical Software}
  \bibinfo{volume}{84}.
\bibitem[{Liu(2012)}]{liu2012some}
\bibinfo{author}{Liu, H.}, \bibinfo{year}{2012}.
\newblock \bibinfo{title}{Some models for time series of counts}.
\newblock \bibinfo{publisher}{Columbia University}.
\bibitem[{Mart{\'\i}nez-Beneito et~al.(2008)Mart{\'\i}nez-Beneito,
  L{\'o}pez-Quilez and Botella-Rocamora}]{martinez2008autoregressive}
\bibinfo{author}{Mart{\'\i}nez-Beneito, M.A.},
  \bibinfo{author}{L{\'o}pez-Quilez, A.}, \bibinfo{author}{Botella-Rocamora,
  P.}, \bibinfo{year}{2008}.
\newblock \bibinfo{title}{An autoregressive approach to spatio-temporal disease
  mapping}.
\newblock \bibinfo{journal}{Statistics in medicine} \bibinfo{volume}{27},
  \bibinfo{pages}{2874--2889}.
\bibitem[{Meyn and Tweedie(2009)}]{meyn2012markov}
\bibinfo{author}{Meyn, S.P.}, \bibinfo{author}{Tweedie, R.L.},
  \bibinfo{year}{2009}.
\newblock \bibinfo{title}{Markov chains and stochastic stability}.
\newblock \bibinfo{publisher}{Cambridge University Press}.
\bibitem[{Mohler(2013)}]{mohler2013modeling}
\bibinfo{author}{Mohler, G.}, \bibinfo{year}{2013}.
\newblock \bibinfo{title}{Modeling and estimation of multi-source clustering in
  crime and security data}.
\newblock \bibinfo{journal}{The Annals of Applied Statistics}
  \bibinfo{volume}{7}, \bibinfo{pages}{1525--1539}.
\bibitem[{Pease et~al.(1998)}]{pease1998repeat}
\bibinfo{author}{Pease, K.}, et~al., \bibinfo{year}{1998}.
\newblock \bibinfo{title}{Repeat victimisation: Taking stock}.
  volume~\bibinfo{volume}{90}.
\newblock \bibinfo{publisher}{Home Office Police Research Group London}.
\bibitem[{Polvi et~al.(1991)Polvi, Looman, Humphries and Pease}]{polvi1991time}
\bibinfo{author}{Polvi, N.}, \bibinfo{author}{Looman, T.},
  \bibinfo{author}{Humphries, C.}, \bibinfo{author}{Pease, K.},
  \bibinfo{year}{1991}.
\newblock \bibinfo{title}{The time course of repeat burglary victimization}.
\newblock \bibinfo{journal}{The British Journal of Criminology}
  \bibinfo{volume}{31}, \bibinfo{pages}{411--414}.
\bibitem[{Polvi et~al.(1990)}]{polvi1990repeat}
\bibinfo{author}{Polvi, N.}, et~al., \bibinfo{year}{1990}.
\newblock \bibinfo{title}{Repeat victimization}.
\newblock \bibinfo{journal}{Journal of Police Science and Administration}
  \bibinfo{volume}{17}, \bibinfo{pages}{8--11}.
\bibitem[{Python et~al.(2019)Python, Illian, Jones-Todd and
  Blangiardo}]{python2019bayesian}
\bibinfo{author}{Python, A.}, \bibinfo{author}{Illian, J.B.},
  \bibinfo{author}{Jones-Todd, C.M.}, \bibinfo{author}{Blangiardo, M.},
  \bibinfo{year}{2019}.
\newblock \bibinfo{title}{A bayesian approach to modelling subnational spatial
  dynamics of worldwide non-state terrorism, 2010--2016}.
\newblock \bibinfo{journal}{Journal of the Royal Statistical Society: Series A
  (Statistics in Society)} \bibinfo{volume}{182}, \bibinfo{pages}{323--344}.
\bibitem[{Raphael and Winter-Ebmer(2001)}]{raphael2001identifying}
\bibinfo{author}{Raphael, S.}, \bibinfo{author}{Winter-Ebmer, R.},
  \bibinfo{year}{2001}.
\newblock \bibinfo{title}{Identifying the effect of unemployment on crime}.
\newblock \bibinfo{journal}{The Journal of Law and Economics}
  \bibinfo{volume}{44}, \bibinfo{pages}{259--283}.
\bibitem[{Reinhart(2018)}]{reinhart2018review}
\bibinfo{author}{Reinhart, A.}, \bibinfo{year}{2018}.
\newblock \bibinfo{title}{A review of self-exciting spatio-temporal point
  processes and their applications}.
\newblock \bibinfo{journal}{Statistical science} \bibinfo{volume}{33},
  \bibinfo{pages}{299--318}.
\bibitem[{Reinhart and Greenhouse(2018)}]{reinhart2018self}
\bibinfo{author}{Reinhart, A.}, \bibinfo{author}{Greenhouse, J.},
  \bibinfo{year}{2018}.
\newblock \bibinfo{title}{Self-exciting point processes with spatial
  covariates: modelling the dynamics of crime}.
\newblock \bibinfo{journal}{Journal of the Royal Statistical Society: Series C
  (Applied Statistics)} \bibinfo{volume}{67}, \bibinfo{pages}{1305--1329}.
\bibitem[{Rodrigues and Assun{\c{c}}{\~a}o(2012)}]{rodrigues2012bayesian}
\bibinfo{author}{Rodrigues, E.C.}, \bibinfo{author}{Assun{\c{c}}{\~a}o, R.},
  \bibinfo{year}{2012}.
\newblock \bibinfo{title}{Bayesian spatial models with a mixture neighborhood
  structure}.
\newblock \bibinfo{journal}{Journal of Multivariate Analysis}
  \bibinfo{volume}{109}, \bibinfo{pages}{88--102}.
\bibitem[{Rue et~al.(2009)Rue, Martino and Chopin}]{rue2009approximate}
\bibinfo{author}{Rue, H.}, \bibinfo{author}{Martino, S.},
  \bibinfo{author}{Chopin, N.}, \bibinfo{year}{2009}.
\newblock \bibinfo{title}{Approximate bayesian inference for latent gaussian
  models by using integrated nested laplace approximations}.
\newblock \bibinfo{journal}{Journal of the royal statistical society: Series b
  (statistical methodology)} \bibinfo{volume}{71}, \bibinfo{pages}{319--392}.
\bibitem[{Short et~al.(2008)Short, D'ORSOGNA, Pasour, Tita, Brantingham,
  Bertozzi and Chayes}]{short2008statistical}
\bibinfo{author}{Short, M.B.}, \bibinfo{author}{D'ORSOGNA, M.R.},
  \bibinfo{author}{Pasour, V.B.}, \bibinfo{author}{Tita, G.E.},
  \bibinfo{author}{Brantingham, P.J.}, \bibinfo{author}{Bertozzi, A.L.},
  \bibinfo{author}{Chayes, L.B.}, \bibinfo{year}{2008}.
\newblock \bibinfo{title}{A statistical model of criminal behavior}.
\newblock \bibinfo{journal}{Mathematical Models and Methods in Applied
  Sciences} \bibinfo{volume}{18}, \bibinfo{pages}{1249--1267}.
\bibitem[{Wall(2004)}]{wall2004close}
\bibinfo{author}{Wall, M.M.}, \bibinfo{year}{2004}.
\newblock \bibinfo{title}{A close look at the spatial structure implied by the
  car and sar models}.
\newblock \bibinfo{journal}{Journal of statistical planning and inference}
  \bibinfo{volume}{121}, \bibinfo{pages}{311--324}.
\bibitem[{Wolpert and Ickstadt(1998)}]{wolpert1998poisson}
\bibinfo{author}{Wolpert, R.L.}, \bibinfo{author}{Ickstadt, K.},
  \bibinfo{year}{1998}.
\newblock \bibinfo{title}{Poisson/gamma random field models for spatial
  statistics}.
\newblock \bibinfo{journal}{Biometrika} \bibinfo{volume}{85},
  \bibinfo{pages}{251--267}.
\bibitem[{Zeevi and Glynn(2004)}]{zeevi2004recurrence}
\bibinfo{author}{Zeevi, A.}, \bibinfo{author}{Glynn, P.W.},
  \bibinfo{year}{2004}.
\newblock \bibinfo{title}{Recurrence properties of autoregressive processes
  with super-heavy-tailed innovations}.
\newblock \bibinfo{journal}{Journal of applied probability}
  \bibinfo{volume}{41}, \bibinfo{pages}{639--653}.
\bibitem[{Zhuang and Mateu(2019)}]{zhuang2019semiparametric}
\bibinfo{author}{Zhuang, J.}, \bibinfo{author}{Mateu, J.},
  \bibinfo{year}{2019}.
\newblock \bibinfo{title}{A semiparametric spatiotemporal hawkes-type point
  process model with periodic background for crime data}.
\newblock \bibinfo{journal}{Journal of the Royal Statistical Society: Series A
  (Statistics in Society)} \bibinfo{volume}{182}, \bibinfo{pages}{919--942}.

\end{thebibliography}






\appendix

\section{Proof of Proposition 1}

We will first make use of the result given in \cite{athreya1986mixing} that states for an AR(1) process, $\lambda=(\lambda_n:n\geq 0)$, $|\eta|<1$ given as
\begin{equation}
\lambda_{n+1}=\eta \lambda_{n}+Z_{n+1},
\end{equation}
where $Z_{n}$ are a sequence of random, i.i.d., variables, a sufficient condition for the existence of a unique stationary distribution is $E[\log(1+|Z_1|)]<\infty$.  This is extended to Vector AR(1) models in \cite{zeevi2004recurrence}.

Due to the temporal independence of $\exp(X(s_i,t)$, we note that the latent process, $\lambda(s_i,t)$, for the SCSE(1,0) process can be written as the VAR(1) process
\begin{equation}
\boldsymbol{\lambda}_t=\exp(\boldsymbol{Y}_t)+\kappa\boldsymbol{\lambda_{t-1}},
\end{equation}
where $\boldsymbol{\lambda}_{t} = (\lambda(s_1,t),\lambda(s_2,t),\cdots,\lambda(s_n,t)^T$.  Here, $\boldsymbol{Y}_t \sim \mbox{iid Gau}\left(\boldsymbol{\alpha}_t,(I-C)^{-1}M\right)$.  Thus, as $E[\log \left(1+||\boldsymbol{Y}_1||\right)]<\infty$ we can appeal to Proposition 2 of \cite{zeevi2004recurrence} and conclude that $\lambda$ admits a unique stationary distribution, $\pi$ and $\lambda$ converges in distribution to $\pi$ as $T\to \infty$.

We next prove geometric ergodicity, and hence stationarity,for $\lambda(s_i,t)$ under a more general formulation

\begin{align}
\lambda(s_i,t)= \exp(X(s_i,t))+\kappa \lambda(s_i,t-1) + \eta Y(s_i,t-1) \label{eq:General}.
\end{align}
First we note that by recursion we can write

\begin{align*}
& [\lambda(s_i,t)|\lambda(s_i,0)=B] = \exp(X(s_i,t))+\kappa \lambda(s_i,t-1) + \eta Y(s_i,t-1)\\
& = \exp(X(s_i,t))+\kappa \left[\exp(X(s_i,t-1))+\kappa \lambda(s_i,t-2) + \eta Y(s_i,t-2)\right] + \eta Y(s_i,t-1)\\
&\cdots\\
& =\sum_{k=0}^{t-1} \kappa\exp(X(s_i,t-k)) +\sum_{k=0}^{t-1} \kappa\eta Y(s_i,t-k-1)+\kappa B. \label{eq:Recursion}
\end{align*}
Intuitively this suggests that the impacts of the initial condition decay at an exponential rate and both the log-Gaussian and the Poisson errors further decay at a geometric rate.  The general proofs of geometric ergodicity for INGARCH properties either follow \cite{davis2016theory} and rely on showing a geometric moment contraction condition, or follow \cite{fokianos2009poisson} and show a drift condition and associated small set condition.  The model given in \eqref{eq:General} cannot easily be shown to satisfy the geometric moment contraction condition as $E[|\exp(\boldsymbol{Y}_{t_i})-\exp(\boldsymbol{Y}_{t_j})|]>0$ for $i \neq j$, so therefore we will closely follow \cite{fokianos2009poisson} and show that a drift condition holds off of a compact set, $C$, then show that $C$ is a small set, see e.g. \cite{meyn2012markov}. 

A precursor to geometric ergodicity is the notion of $\phi$-irreducibility.  $\phi$-irreducbility, as defined in \cite{meyn2012markov}, formally is that there exists a measure, $\phi$, such that for all Borel sets, $A$, such that $\phi(A)>0$, $\mbox{P}_{\lambda_0}(\tau_A<\infty)>0$, where $\mbox{P}_{\lambda_0}$ is the Markov chain beginning at $\lambda_0$ and $\tau_A$ is the hitting time of the Markov chain.  In other words, $\phi$-irreducibility means that for any set that has positive measure, the Markov chain has a positive probability of eventually entering the set.

$\phi$-irreducible further implies and is implied by the condition that for every set $A$ with $\phi(A)>0$, $P(\lambda(s_i,t)\in A|\lambda(s_i,0)=B)>0$ for some $t$.  
To demonstrate this, consider the sequence, $Y(s_i,0)=Y(s_i,1)=\cdots=Y(s_i,t-1)=0$ which occurs with positive probability due to the conditional Poisson density of $Y(s_i,t)$ assuming $\lambda(s_i,0)<\infty$.  If we choose $t$ large enough we can always have $\kappa\lambda(s_i,0) < \mbox{inf } A$, hence along this sequence $P(\lambda(s_i,t)\in A|\lambda(s_i,0)=B)= P[\sum_{k=0}^{t-1} \kappa\exp(X(s_i,t-k))\in (A-\kappa^t B)]$.  Though there is no close formed density for sums of log-normals, $\sum_{k=0}^{t-1} \kappa^k\exp(X(s_i,t-k))$ clearly has positive measure on $\mathbb{R}^{+}$.  Therefore for any $A$ with $\phi(A)>0$, $P[\lambda(s_i,t)\in A|\lambda(s_i,0)=B]>P[\lambda(s_i,t)\in A|\lambda(s_i,0)=B,Y(s_i,0)=Y(s_i,1)=\cdots=Y(s_i,t-1)=0]P[Y(s_i,0)=Y(s_i,1)=\cdots=Y(s_i,t-1)=0]|\lambda(s_i,0)=B>0$ which implies $\phi$ irreducibility.  Note that a similar argument gives aperiodicity.

Next, we will appeal to Theorem 15.0.1 (iii) and Lemma 15.2.8 of \cite{meyn2012markov} in a manner similar to \cite{fokianos2009poisson}. We will first show that there exists a test function, $V(\lambda)$ where the inequality $E[V(\boldsymbol{\lambda}_{t+1})|\boldsymbol{\lambda}_t=\boldsymbol{\lambda_*}]\leq \psi V(\boldsymbol{\lambda_*})+L \mbox{ I}(\boldsymbol{\lambda_*} \in C)$ holds where $\psi \in (0,1)$, $L \in (0,\infty)$ and $I(.)$ is the indicator function.  Next we will show that $C$ is a small set and hence a petite set.

To simplify notation we consider spatial $n=1$ so our compact set $C$ is in $\mathbb{R}$ rather then the multi-dimensional case for a general spatial region, though everything that follows can be extended to $n\in\mathbb{N}^+$.  The compact set we will consider is $C \in [0,G]$ where  $G \in (0,\infty)$.  Note that this requires expanding the parameter space of $\lambda$ from $(0,\infty)$ to $[0,\infty)$.

Akin to \cite{fokianos2009poisson} we consider $V(\lambda)=1+\lambda^2$.  Suppressing the dependency on $\alpha, \zeta,\mbox{ and },\sigma_{sp}^2$ we write $\gamma$ as the second moment of the log-Gaussian density, $\exp(X(s_i,t))$ and have

\begin{align}
& E[V(\lambda_t)|\lambda_{t-1}=\lambda_*]= 1+E[\left(\exp(X(s_i,t))+\kappa\lambda_{t-1}+\eta Z_{t-1}\right)|\lambda_{t-1}=\lambda_*]\\
& = 1+\gamma+(\kappa+\eta)^2\lambda_*^2+2(\eta+\kappa)\exp(\alpha+\frac{\Sigma_{1,1}}{2})\lambda_* \label{eq:V}
\end{align}
First consider $\lambda_* \in C^c$, on this set we have

\begin{align}
& 1+\gamma+(\kappa+\eta)^2\lambda_*^2+2(\eta+\kappa)\exp(\alpha+\frac{\Sigma_{1,1}}{2})\lambda_* =\nonumber\\ & \left[\left(1-\frac{\lambda_*^2}{1+\lambda_*^2}+\frac{(\eta+\kappa)^2 \lambda_*^2}{1+\lambda_*^2}\right) + \frac{\gamma}{1+\lambda_*^2}+\frac{2(\eta+\kappa)\exp(\alpha+\frac{\Sigma_{1,1}}{2})\lambda_*}{1+\lambda_*^2}\right](1+\lambda_*^2).
\end{align}
Here, as $G$ increases, the supremum of the term inside $\left[.\right]$ goes to $(\eta+\kappa)^2$ which is less than 1 by assumptions on the parameter space for $\eta$ and $\kappa$.  

Next, for $\lambda_* \in C$, we can still write \eqref{eq:V} which is bounded by $1+\gamma+(\kappa+\eta)^2 G^2 + 2(\eta+\kappa)\exp(\alpha+\frac{\Sigma_{1,1}}{2})G$.  Thus, there exists $C=[0,G]$ such that $E[V(\lambda_{t+1})|\lambda_t=\lambda_*]\leq \psi V(\lambda_*)+L \mbox{ I}(\lambda_*\in C)$.

Next, we will show that the set $C=[0,G]$ is a small set.  That is, $\exists \mbox{ }n$ such that
\begin{equation}
\inf_{\lambda \in C}\mbox{P}^n(\lambda,A)>0
\end{equation}
for a set $A$ having Lebesgue measure greater than zero.

To show this, let $\lambda_0 \in C$ and $Z_0=0$.  Then, there exists a path, $Z_1=\cdots=Z_m=0$ that exists with probability greater than zero.  Using the recursive formulation, \eqref{eq:Recursion}, it follows that along that path, $\lambda_m=\sum_{k=0}^{m-1}\kappa\exp(Y_{t-k})+(\kappa)^m \lambda_0$.  While the geometric sum of uncorrelated log Gaussian terms, $\sum_{k=0}^{m-1}\kappa\exp(Y_{t-k})$ has no closed form solution, it has density with regard to the positive Lebesgue measure.  Therefore, if we consider an interval with positive Lebesgue measure, $A=(a,b)$, then there exists $m=N$ such that $\kappa^N \lambda_0<a$.  Therefore, for $A$, $\inf_{\lambda\in C}  \mbox{P}^N(\lambda,A) > \mbox{P }(\kappa\exp(Y_{t-k}) \in (a-\kappa)^N \lambda_0,b-(\kappa)^N \lambda_0)\mbox{P }(Z_1=\cdots=Z_N=0)>0$.  Thus, the interval $(a,b)$ is uniformly reachable from $\lambda_0 \in C$.  Therefore it follows in a manner similar to \cite{fokianos2009poisson}, that $C=[0,K]$ is a small set.  This demonstrates that \eqref{eq:General} is geometrically ergodic and therefore admits a unique stationary distribution.  Furthermore, the specific choice of $V(.)$ used in the drift condition ensures that second moments exist for the stationary distribution.

\section{Derivation of Second Order Properties}

\subsection{Derivation of Variance}
To see how the variance to mean ratio can be adjusted under the SCSE model we can first compute the marginal variance of $Y(\boldsymbol{s_i},t)$. To find this value we exploit the independence of $U(\si,t)\equiv X(\si)+\epsilon(\si,t)$ and $Y(\boldsymbol{s_i},t-1)$ yielding

\begin{align}
	\mbox{Var}(Y(\boldsymbol{s_i},t+1))  =& \mbox{Var}(E(Y(\boldsymbol{s_i},t+1)|\lambda(\boldsymbol{s_i},t+1))+E(\mbox{Var}(Y(\boldsymbol{s_i},t+1)|\lambda(\boldsymbol{s_i},t+1))\\
	=& \mbox{Var}(\lambda(\boldsymbol{s_i},t+1))+ E(\lambda(\boldsymbol{s_i},t+1))\nonumber\\
	= & \kappa^2 \mbox{Var} (\lambda(\boldsymbol{s_i},t))+\eta^2 \mbox{Var}(Y(\boldsymbol{s_i},t))+2 \kappa \eta \mbox{Var }\lambda(\boldsymbol{s_i},t)+\nonumber\\
	& \mbox{Var}\left[\exp(U(\boldsymbol{s_i},t))\right]+E(Y(\boldsymbol{s_i},t))\\
	=&\kappa^2 \mbox{Var} (Y(\boldsymbol{s_i},t))+\eta^2 \mbox{Var}(Y(\boldsymbol{s_i},t))+2 \kappa \eta \mbox{Var }(Y(\boldsymbol{s_i},t))+\nonumber \nonumber\\
	& - \kappa^2 E(Y(\boldsymbol{s_i},t))-2 \kappa \eta E(Y(\boldsymbol{s_i},t))+\mbox{Var} \left[\exp(U(\boldsymbol{s_i},t))\right].
\end{align}

\subsection{Derivation of Temporal and Spatial Covariance}

In order to derive the temporal covariance, without loss of generality we assume $\boldsymbol{\alpha}=\boldsymbol{0}$ and we first find

\begin{align}
& E\left[Y(s_i,t)Y(s_i,t+1)\right] = E\left[Y(s_i,t)\left(E[Y(s_i,t+1)|\lambda(s_i,t),Y(s_i,t)\right])\right]\\
& = E\left[Y(s_i,t)\left(\eta Y(s_i,t)+\kappa \lambda(s_i,t)+\exp(\frac{\Sigma_{1,1}}{2})\right)\right]\\
& = E\left[Y(s_i,t)^2\right]+\kappa E\left[Y(s_i,t)\lambda(s_i,t)\right]+\frac{1}{1-(\eta+\kappa)}\exp(\Sigma_{1,1})\\
& = \eta \left(\mbox{Var }(Y(s_i,t))+E[Y(s_i,t)]^2\right)+\kappa E[\lambda(s_i,t)^2]+\frac{1}{1-(\eta+\kappa)}\exp(\Sigma_{1,1})\\
& = \eta \mbox{Var }(Y(s_i,t))+\frac{\eta}{(1-(\eta+\kappa))^2}\exp(\Sigma_{1,1})+\kappa E[\lambda(s_i,t)^2]+\frac{1-(\eta+\kappa)}{(1-(\eta+\kappa))^2}\exp(\Sigma_{1,1}).
\end{align}
Therefore, as $E[Y(s_i,t)]^2=\frac{1}{(1-(\eta+\kappa))^2}\exp(\Sigma_{1,1})$, the covariance is

\begin{align}
\mbox{Cov }(Y(s_i,t)Y(s_i,t+1))=\eta \mbox{Var}Y(s_i,t)+\kappa E[\lambda(s_i,t)^2]-\frac{\kappa}{(1-(\eta+\kappa))^2}\exp(\Sigma_{1,1}).
\end{align}
Next, we note that $E[\lambda(s_i,t)^2]=\mbox{Var }(\lambda(s_i,t))+\left(E[\lambda(s_i,t)]\right)^2=\mbox{Var}(Y(s_i,t))-E[\lambda(s_i,t)]+\left(E[\lambda(s_i,t)]\right)^2$.  Thus we have

\begin{align}
& \mbox{Cov }(Y(s_i,t)Y(s_i,t+1))=\left(\eta+\kappa\right)\mbox{Var}(Y(s_i,t))-\kappa E[\lambda(s_i,t)]+\nonumber\\
&\frac{\kappa}{(1-(\eta+\kappa))^2}\exp(\Sigma_{1,1})-\frac{\kappa}{(1-(\eta+\kappa))^2}\exp(\Sigma_{1,1})\\
&=\left(\eta+\kappa\right)\mbox{Var}(Y(s_i,t))-\kappa E[Y(s_i,t)],
\end{align}
as desired.

Next we find $E[Y(s_i,t)Y(s_j,t)]$ for arbitrary $i \neq j$.  Recall that we let $\Sigma_{i,j}$ be the entry in the covariance matrix at location $(i,j)$.

First note that $E[Y(s_i,t)Y(s_j,t)]=E[E[Y(s_i,t)Y(s_j,t)|\lambda(s_i,t),\lambda(s_j,t)]]=E[\lambda(s_i,t)\lambda(s_j,t)]$.  Using this we have

\begin{align}
& E[Y(s_i,t)Y(s_j,t)]=E[\lambda(s_i,t)\lambda(s_j,t)]\\
& = \eta^2 E[Y(s_i,t-1)Y(s_j,t-1)] + \eta \kappa E[Y(s_i,t-1)\lambda(s_j,t-1)]+\nonumber\\
& \eta \kappa E[\lambda(s_i,t-1)Y(s_j,t-1)]+2\eta \frac{1}{1-(\eta+\kappa)}\exp(2\alpha \Sigma_{i,i})+\nonumber \\
&\kappa^2 E[\lambda(s_i,t-1)\lambda(s_j,t-1)]+2 \kappa \frac{1}{1-(\eta+\kappa)}\exp(2\alpha \Sigma_{i,i})+\exp(2\alpha)\exp(\Sigma_{i,i}+\Sigma_{i,j})\\
& = (\eta+\kappa)^2 E[Y(s_i,t-1)Y(s_j,t-1)]+2\frac{\eta+\kappa}{1-(\eta+\kappa)}\exp(\Sigma_{i,i})+\exp(2\alpha)\exp(\Sigma_{i,i}+\Sigma_{i,j})
\end{align}
Relying on second order stationarity in time, this yields

\begin{align}
& E[Y(s_i,t)Y(s_j,t)] = \frac{1}{1-(\eta+\kappa)^2}\left(2\frac{(\eta+\kappa)}{1-(\eta+\kappa)}\exp(2\alpha+\Sigma_{i,i})+\exp(\Sigma_{i,i}+\Sigma_{i,j})\right).
\end{align}
Therefore, we have

\begin{align}
& \mbox{Cov} (Y(s_i,t)Y(s_j,t)=\frac{1}{1-(\eta+\kappa)^2}\left(2\frac{(\eta+\kappa)}{1-(\eta+\kappa)}\exp(2\alpha+\Sigma_{i,i})+\exp(\Sigma_{i,i}+\Sigma_{i,j})\right) -\nonumber\\ & \frac{1}{(1-(\eta+\kappa))^2}\exp(2\alpha + \Sigma_{i,i})\\
&=\frac{\exp(2\alpha)}{1-(\eta+\kappa)^2}\left(\exp(\Sigma_{i,i}+\Sigma_{i,j}-\exp(\Sigma_{i,i}))\right),
\end{align}
as given in \eqref{eq:SpatCov}. 

\section{Relationship between SCSE Model and Self-Exciting Processes}\label{sec:SCSE and Self-Exciting Process}

In order to motivate the model we begin with a point process similar to a Hawkes process, \cite{hawkes1971spectra}, as described in \cite{laub2015hawkes}.  The Hawkes process is a temporal process where $\lambda(t)=\alpha + \int_0^t g(t-u) dN(u)$, here integration is taken over some counting process.  Extensions can be made to spatio-temporal processes, as in \cite{reinhart2018review} through modeling the intensity function at space-time location $(\boldsymbol{s_i},t)$ as

\begin{equation}
     \lam= \alpha (\boldsymbol{s_i},t)+\sum_{\{j:t_j<t}f(|\si-\boldsymbol{s_j}|)h(t-t_j)\label{eq:sthawkes}.
\end{equation}

The mechanistic interpretation for the Hawkes process is straight forward.  There is a spatially or temporally varying background intensity $\alpha(\si,t)$ as well as an self-excitement term, $\sum_{\{j:t_j<t}f(\si-\boldsymbol{s_j})h(t-t_j)$ which allows for a positive feedback from observed events that occur from a spatial distance of $|s_i - s_j|$ from previously observed sequences of locations of events.  Here the $f(.)$ function accounts for the distance between $s_i$ and the previous events whereas $h(t-t_j)$ accounts for the time between $t$ and previous events. The summation occurs over all discrete time periods.  Commonly, the process is simplified to only allow temporal excitement and uses an exponential kernel for $h(t-t_j)=\eta\exp(-\beta(t-t_j))$ \eqref{eq:sthawkes} thus becomes,

\begin{equation}
     \lam= \alpha (\boldsymbol{s_i},t)+\sum_{j:t_j<t}\eta\exp(-\beta(t-t_j))\label{eq:hawkesexp}.
\end{equation}

Once an event occurs, say at time $t_j$, $\beta$ controls how long that impacts the intensity of the process while $\eta$ impacts the amount that the event impacts $\lam$.

If, however, our data is presented aggregated over time, then for each $t_j$, we have $0,1,2,\ldots$ events. Therefore, if $Y(\si,t)$ is the number of occurrences that happen at $\si$ between $t$ and $t-1$  \eqref{eq:hawkesexp} becomes

\begin{equation}
     \lam= \alpha (\boldsymbol{s_i},t)+\sum_{j=1}^n Y(\si,t-j)\eta\exp(-\beta j)\label{eq:hawkesdisctime}.
\end{equation}

To demonstrate how this relates to a common discrete-valued time series we re-parameterize $\kappa=\exp(-\beta)$ and $\alpha(\si,t)=\frac{1-\kappa^{n-1}}{1-\kappa}U (\si,t)$. Note that this is a one to one reparameterization that changes the interpretation of the parameters without impacting the random variables in the model. The reparameterization yields

\begin{equation}
     \lam= \frac{1-\kappa^{n-1}}{1-\kappa}U (\si,t)+\sum_{j=1}^n Y(\si,t-j)\eta\kappa^j \label{eq:almostingarch},
\end{equation}

which can recursively be re-written as

\begin{equation}
    \lam=U(\si,t) + \kappa \lambda(\si,t-1) + \eta Y(\si,t-1)\label{eq:ingarch}.
\end{equation}
Which we note the parallels to the Poisson auto-regression model of \cite{fokianos2009poisson}.  

To ensure we obey the parameter space for \lam  we restrict $U(\si,t)>0$, to do this we allow $U(\si,t) \equiv\exp\left(X(\si)+\epsilon(\si,t)\right)$. Note here we assume the latent process can be decomposed into a spatially varying component, $X(\si)$ and a spatially temporally varying component $\epsilon(\si,t)$.

\end{document}